\documentclass[pdflatex,sn-apa]{sn-jnl}

\usepackage{graphicx}%
\usepackage{multirow}%
\usepackage{amsmath,amssymb,amsfonts}%
\usepackage{amsthm}%
\usepackage{mathrsfs}%
\usepackage[title]{appendix}%
\usepackage{xcolor}%
\usepackage{textcomp}%
\usepackage{manyfoot}%
\usepackage{booktabs}%
\usepackage{algorithm}%
\usepackage{algorithmicx}%
\usepackage{algpseudocode}%
\usepackage{listings}%

\newtheorem{lemma}{Lemma}
\newtheorem{corollary}{Corollary}
\newtheorem{observation}{Observation}
\usepackage{adjustbox}
\usepackage{diagbox}
\usepackage{bm}
\PassOptionsToPackage{authoryear}{natbib}
\usepackage{natbib}
\usepackage{float}
\usepackage{chngcntr}
\usepackage{subcaption}
\usepackage{siunitx}

\theoremstyle{thmstyleone}%
\newtheorem{theorem}{Theorem}
\newtheorem{proposition}[theorem]{Proposition}%

\theoremstyle{thmstyletwo}%
\newtheorem{remark}{Remark}%

\theoremstyle{thmstylethree}%
\newtheorem{definition}{Definition}%

\begin{document}

\title[Article Title]{Toward Optimizing the Expected Performance of Sampling-Based Quantum-Inspired Algorithms}



\author[1]{\fnm{Hyunho} \sur{Cha}}\email{aiden132435@cml.snu.ac.kr}
\author[1]{\fnm{Sunbeom} \sur{Jeong}}\email{sb3991@cml.snu.ac.kr}
\author*[1]{\fnm{Jungwoo} \sur{Lee}}\email{junglee@snu.ac.kr}
\affil[1]{\orgdiv{NextQuantum and Department of Electrical and Computer Engineering}, \orgname{Seoul National University}, \orgaddress{\city{Seoul}, \postcode{08826}, \country{Republic of Korea}}}


\abstract{Quantum-inspired classical algorithms has received much attention due to its exponential speedup compared to existing algorithms, under certain data storage assumptions. The improvements are noticeable in fundamental linear algebra tasks. In this work, we analyze two major subroutines in sampling-based quantum-inspired algorithms—specifically, inner product estimation and sampling from a linear combination of vectors—and discuss their possible improvements by generalizing the data structure. The idea is to consider the average behavior of the subroutines under certain assumptions regarding the data elements. This allows us to determine the optimal data structure, and the high-dimensional nature of data makes our assumptions reasonable. Experimental results from recommendation systems also highlight a consistent preference for our proposed data structure. Motivated by this observation, we tighten the upper bound on the number of required measurements for direct fidelity estimation. We expect our findings to suggest optimal implementations for various quantum and quantum-inspired machine learning algorithms that involve extremely high-dimensional operations, which has potential for many applications.}




\maketitle

\section*{Acknowledgments}
This work is in part supported by the National Research Foundation of Korea (NRF, RS-2024-00451435 (20\%), RS-2024-00413957 (40\%)), Institute of Information \& Communications Technology Planning \& Evaluation (IITP, 2021-0-01059 (40\%)), grant funded by the Ministry of Science and ICT (MSIT), Institute of New Media and Communications (INMAC), and the Brain Korea 21 FOUR program of the Education and Research Program for Future ICT Pioneers.

\section{Introduction}
Quantum algorithms have intrigued researchers for years with the potential of achieving exponential speedup compared to classical algorithms \citep{deutsch1992rapid, shor1999polynomial, ronnow2014defining, montanaro2015quantum, kerenidis2016quantum, childs2018toward}. These algorithms leverage the principles of quantum mechanics to perform computations in parallel, potentially solving certain problems much more efficiently than classical algorithms ever could.

However, contrary to prevailing belief, it was recently shown that under certain conditions—when data is structured in a particular way—some quantum machine learning (QML) algorithms do not exhibit exponential speedup over classical ones \citep{gilyen2018quantum, tang2019quantum, tang2021quantum, chia2022sampling}. In other words, the exponential speedup of quantum algorithms relied on the assumption that preparing the quantum state encoding of a given classical input could be achieved in sublinear time relative to the input size. This fact was first demonstrated by explicitly constructing a classical recommendation algorithm inspired by its quantum counterpart \citep{tang2019quantum}. If classical data is preprocessed in a proper manner, a dequantized QML algorithm can be designed to run only polynomially slower than the quantum version. Specifically, the data must be prepared in a tree structure to enable efficient sampling comparable to quantum algorithms.

Since the introduction of the quantum-inspired recommendation system, several quantum algorithms have been successfully dequantized \citep{gilyen2018quantum, tang2021quantum, chia2022sampling}, highlighting both the limitation and potential of quantum computing. Under certain conditions, these algorithms are expected to achieve much better performance than known theoretical bounds, especially in extremely high-dimensional problems \citep{arrazola2020quantum}. Therefore, it is instructive to characterize the best possible performance of these quantum-inspired algorithms for precise comparison and further development.

Quantum-inspired algorithms achieve exponential speedup by efficiently dequantizing two important subroutines, which are major computational bottlenecks in many quantum-inspired algorithms: \textit{inner product estimation} and \textit{sampling from a linear combination of vectors}. Here, we assume that data is stored as a binary search tree (BST), with each leaf node storing the square of a data element and each parent node storing the sum of the values in its child nodes. This structure enables efficient $L^2$ norm sampling from the probability distribution defined by the vector. Consequently, we can estimate the inner product of two vectors and draw a sample from the distribution defined by a linear combination of vectors in sublinear time relative to the data size.

Despite the resemblance to quantum algorithms, which theoretically provide exponential advantages, these algorithms are purely classical, offering us more freedom in how they are executed. In the context of quantum analogies, the use of an $L^2$ norm data structure is not arbitrary, as probabilities are inherently proportional to the squared amplitudes. However, from a classical perspective, a more direct way to associate probabilities with data elements is to make them proportional to their absolute values, corresponding to an $L^1$ norm data structure. Empirical evidence from a recommendation system example supports this choice. Moreover, if we make further assumptions on the data, then averaging over the data elements allows for a clear expression of each subroutine's performance. We can generalize the data structure to arbitrary norms and analyze the average runtime. Informally stated,
\begin{itemize}
\item \textit{$L^1$ norm data structure is desirable for sampling-based inner product estimation. It minimizes the average variance in estimating the inner product (averaged over the vector elements) under certain assumptions.}
\item \textit{$L^1$ norm data structure is desirable for sampling from a linear combination of vectors. For sparse vectors, it is $\approx O(n)$ times faster than $L^2$ norm sampling, where $n$ is the number of vectors involved in the linear combination. For dense vectors, it is $O(\sqrt{n})$ times faster than $L^2$ norm sampling on average (averaged over the vector elements and the coefficients used in linear combinations) under certain assumptions.}
\end{itemize}
For high-dimensional operations, the law of large numbers and concentration phenomena imply that deviations of the algorithm's global behavior from theoretical predictions are likely to be negligible. As a result, it is claimed that one need to consider the optimal norm when dequantizing a quantum algorithm at hand, as it may vary depending on the algorithm. This aspect is discussed in detail in Section~\ref{section:discussion}.

\begin{figure}[t]
  \centering
  \includegraphics[width=0.7\linewidth]{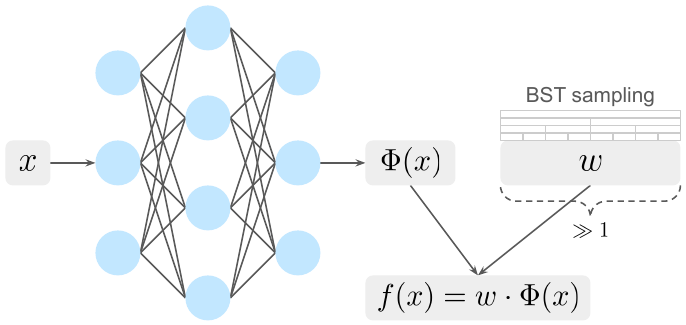}
  \vspace{1em}
  \caption{Input data $x$ is encoded as $\Phi(x)$, with the assumption that querying the elements of $\Phi(x)$ can be done efficiently (in sublinear time with respect to its length). The weight vector $w$ is stored as a BST structure supporting efficient sampling and updating operations. The goal is to efficiently estimate the inner product $f(x) = w \cdot \Phi(x)$ to a desired accuracy.}
  \label{fig:Fig1}
\end{figure}

A well-known scenario where efficient inner product estimation shows its potential is nonlinear kernel learning, where the decision function $f(x) = w \cdot \Phi(x)$ is the inner product of $w$ and $\Phi(x)$ with extremely large dimension (see Figure~\ref{fig:Fig1}). Optimizing the implementation of this inner product estimation is important for numerous applications, as many computational models are expected to work with increasingly higher-dimensional feature spaces in the future. For example, with the advancement of large language models and attention mechanisms operating in higher-dimensional feature spaces, inner product computation may become a key computational bottleneck.

Direct Fidelity Estimation (DFE) \citep{flammia2011direct}, a technique developed in advance of quantum-inspired algorithms, is another practical application of sampling-based inner product estimation. In Section~\ref{sec:dfe_application}, we show that for some target states, we can tighten the upper bound on the number of required measurements to perform DFE by sampling a Pauli operator based on the $L^1$ norm distribution for the characteristic function.

The remainder of this paper is organized as follows. Section~\ref{sec:background} provides an overview of some essential subroutines in quantum-inspired algorithms and the data structures used to implement them. Section~\ref{sec:optimal_norm} presents our main results and analyses of the optimal norm under certain assumptions. Numerical simulation and DFE application are covered in sections~\ref{sec:experimental_results} \& \ref{sec:dfe_application}, respectively. Finally, the discussion and conclusion are presented in sections \ref{section:discussion} \& \ref{sec:conclusion}.

\section{Background}
\label{sec:background}

\subsection{Nomenclature}
Consider $x \in \mathbb{R}^n$ and $A \in \mathbb{R}^{m \times n}$. $||x||$ and $||x||_p$ are the $L^2$ norm and $L^p$ norm of $x$, respectively. $||A||_F$ is the Frobenius norm of $A$. $A^{(j)}$ is the $j$-th column of $A$. $\Tilde{A}$ is a vector in $\mathbb{R}^n$ such that $\Tilde{A}_j = ||A^{(j)}||$ and $\Tilde{A}^{(p)}$ is a vector in $\mathbb{R}^n$ such that $\Tilde{A}^{(p)}_j = ||A^{(j)}||_p$. $\text{nnz}(x)$ ($\text{nnz}(A)$) is the number of nonzero elements in $x$ ($A$). $\mathcal{U}_{(a, b)}$ is the continuous uniform distribution over $(a, b)$. $[n]$ denotes the set $\{ 1, 2, \ldots , n \}$.

\subsection{$\text{SQ}_2$ data structure}
\label{sec:l2_structure}

\begin{definition}
\label{def:l2_distribution}
For a nonzero vector $x \in \mathbb{R}^n$, we denote by $\mathcal{D}_x$ the distribution over $[n]$ with PDF
\begin{equation*}
\mathcal{D}_x(i) = \frac{x_i^2}{||x||^2}.
\end{equation*}
\end{definition}

\noindent Hereafter, the notation $\mathcal{D}_x$ implies that $x$ is a nonzero vector. Definition~\ref{def:l2_distribution} has a quantum analog, since $\mathcal{D}_x$ is the distribution corresponding to the \textit{amplitude encoding} of $x$ \citep{schuld2018supervised}:
\begin{equation*}
|x\rangle = \sum_{i = 1}^{n} x_i |i\rangle.
\end{equation*}

\begin{lemma}[{\citeauthor{tang2019quantum}, \citeyear{tang2019quantum}}]
\label{lemma:vector_structure_exists}
There exists a data structure that stores a vector $x \in \mathbb{R}^n$ in $O(n)$ space (or $O(\textnormal{nnz}(x) \log n)$ for sparse $x$), supporting the following operations:
\begin{itemize}
\item Sample $i \sim \mathcal{D}_x$ in $O(\log n)$ time.
\item Update $x_i$ and the corresponding data structure in $O(\log n)$ time.
\end{itemize}
\end{lemma}

\begin{corollary}[{\citeauthor{tang2019quantum}, \citeyear{tang2019quantum}}]
\label{corollary:matrix_structure_exists}
There exists a data structure that stores a matrix $A \in \mathbb{R}^{m \times n}$ in $O(mn)$ space (or $O(\textnormal{nnz}(A) \log (mn))$ for sparse $A$), supporting the following operations:
\begin{itemize}
\item Sample $j \sim \mathcal{D}_{\Tilde{A}}$ in $O(\log n)$ time.
\item Sample $i \sim \mathcal{D}_{A^{(j)}}$ in $O(\log m)$ time.
\item Update $A_{ij}$ and the corresponding data structure in $O(\log (mn))$ time.
\end{itemize}
\end{corollary}

We may assume that each \textit{query} operation for a quantity stored in the data structure takes $O(1)$ time. For simplicity, the time complexities of algorithms that employ such a data structure can be expressed in terms of query and sample operations.

\begin{figure}[t]
  \centering
  \begin{subfigure}[t]{0.57\linewidth}
    \centering
    \includegraphics[width=\linewidth]{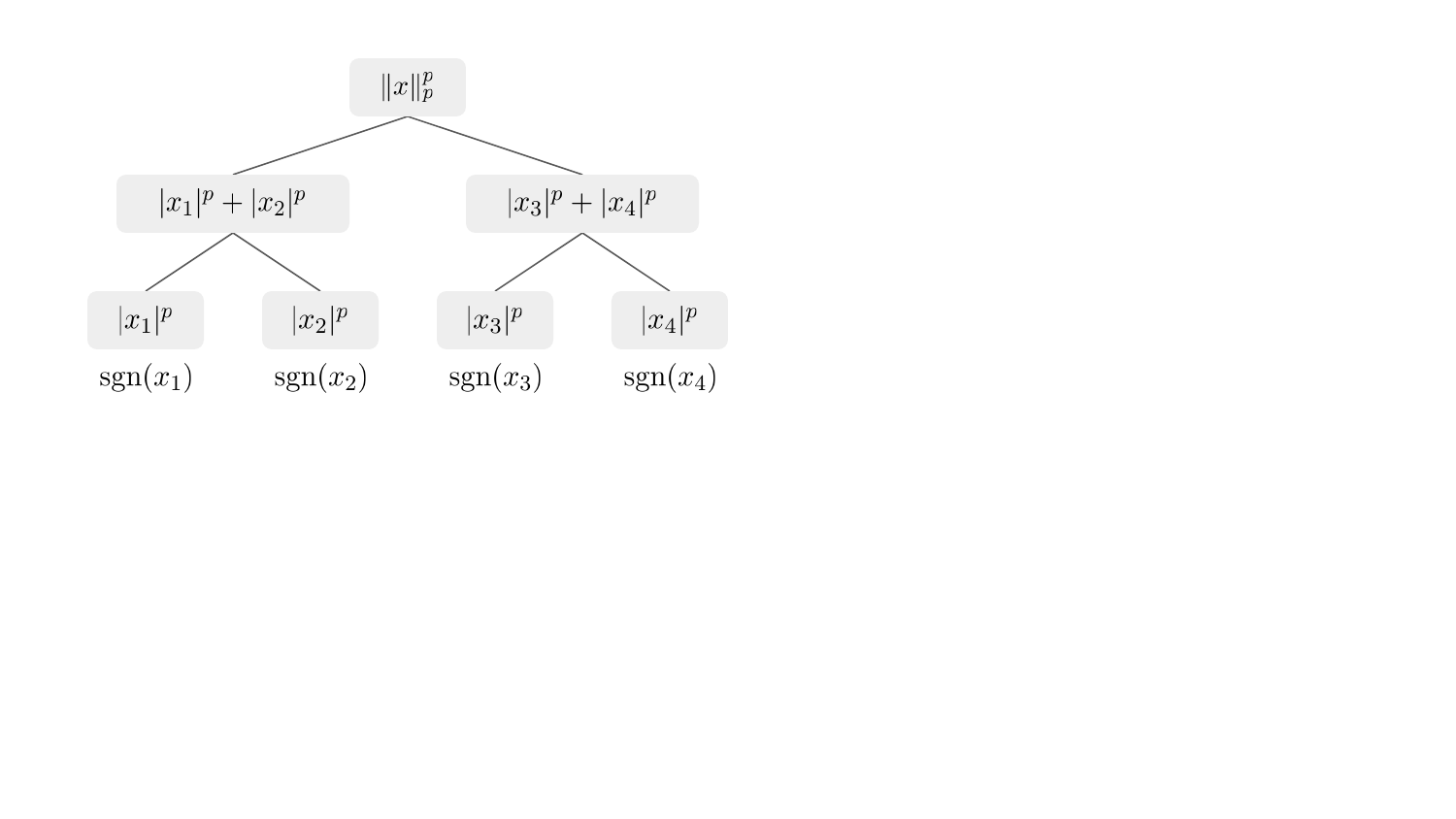}
    \caption{\vphantom{a}}
  \end{subfigure}
  \hspace{0.01\linewidth}
  \begin{subfigure}[t]{0.4\linewidth}
    \centering
    \includegraphics[width=\linewidth]{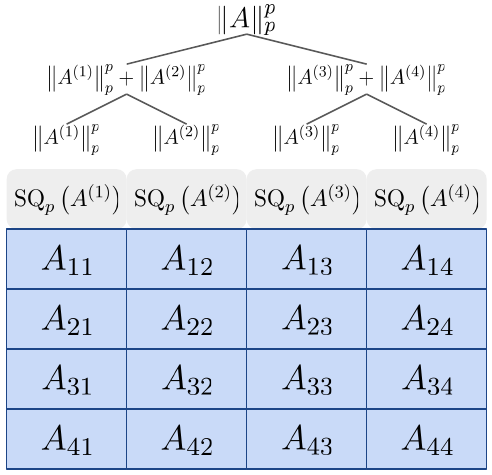}
    \caption{\vphantom{a}}
  \end{subfigure}
  \caption{BST data structures for (a) $x \in \mathbb{R}^4$ (\(\mathrm{SQ}_p(x)\)) and (b) $A \in \mathbb{R}^{4 \times 4}$ (\(\mathrm{SQ}_p(A)\)) based on $L^p$ norm.}
  \label{fig:lp_diagram}
\end{figure}

Figure~\ref{fig:lp_diagram} with $p = 2$ is the data structure that verifies Lemma~\ref{lemma:vector_structure_exists} and Corollary~\ref{corollary:matrix_structure_exists}. They are essentially BST data structures, where the leaf nodes store $x_i^2$ ($A_{ij}^2$) and their sign ($\text{sgn}(x_i)$ or $\text{sgn}(A_{ij})$). Each internal node stores the sum of the values of its child nodes. To draw a single sample from $\mathcal{D}_x$, we start from the root node and traverse down the tree, where each child is selected with probability proportional to the value stored in it. The same technique is applied to draw a sample from a matrix $A$, where we first sample a column $j \sim \mathcal{D}_{\Tilde{A}}$ and then sample a row $i \sim \mathcal{D}_{A^{(j)}}$.

The data structures for Lemma~\ref{lemma:vector_structure_exists} and Corollary~\ref{corollary:matrix_structure_exists}, which allow both query and sample accesses, are referred to as $\text{SQ}_2(x)$ and $\text{SQ}_2(A)$, respectively. If only query access is allowed for a vector $x$, the data structure is denoted as $\text{Q}(x)$.

\subsection{Inner product estimation with $\text{SQ}_2$}
\label{sec:l2_inner_product}

Assuming that $x \in \mathbb{R}^n$ is stored in the data structure described in Section~\ref{sec:l2_structure}, we can efficiently estimate the inner product of $x$ and another vector.

\begin{proposition}[{\citeauthor{tang2019quantum}, \citeyear{tang2019quantum}}]
\label{prop:l2_inner_product_prop}
Given $\textnormal{SQ}_2(x)$ and $\textnormal{Q}(y)$ for $x, y \in \mathbb{R}^n$, the inner product $\langle x, y \rangle$ can be estimated to additive error $\epsilon ||x|| \cdot ||y||$ with at least $1 - \delta$ probability using $O\left(\frac{1}{\epsilon^2} 
\log \frac{1}{\delta} \right)$ queries and samples.
\end{proposition}

\subsection{Sampling from a linear combination of vectors with $\text{SQ}_2$}
\label{sec:l2_combination}

Suppose we are given $\text{SQ}_2(A)$ for $A \in \mathbb{R}^{m \times n}$ and $\text{Q}(x)$ for $x \in \mathbb{R}^n$. Then we can efficiently sample from a linear combination of the vectors $\{ A^{(j)} \}_{j = 1}^{n}$, weighted by $\{ x_j \}_{j = 1}^{n}$. Formally, the following proposition holds.

\begin{proposition}[{\citeauthor{tang2019quantum}, \citeyear{tang2019quantum}}]
\label{prop:l2_linear_combination}
Given $\textnormal{SQ}_2(A)$ for $A \in \mathbb{R}^{m \times n}$ and $\textnormal{Q}(x)$ for $x \in \mathbb{R}^n$, we can output a sample from $\mathcal{D}_{Ax}$ in $O(n^2 C(A, x))$ expected query complexity, where
\begin{equation*}
C(A, x) := \frac{\sum_{j = 1}^{n} ||x_j A^{(j)}||^2}{||Ax||^2}.
\end{equation*}
\end{proposition}

Proposition~\ref{prop:l2_linear_combination} was developed to implement the final step of the quantum-inspired recommendation algorithm, specifically the recommendation phase. The algorithm in the recommendation context assigns items to users based on linear combinations of other relevant users.

\section{Optimizing sample access}
\label{sec:optimal_norm}
The subroutines introduced in sections \ref{sec:l2_inner_product} and \ref{sec:l2_combination} are essential for a wide range of machine learning tasks. Therefore, optimizing these tasks will enhance the potential of quantum-inspired machine learning algorithms that utilize these steps.

The $L^2$ norm sample access arises naturally from the quantum encoding of classical vectors, which is widely utilized in the development of quantum algorithms \citep{rebentrost2014quantum, kerenidis2016quantum, park2020theory, schuld2020circuit}. For a classical vector $x \in \mathbb{R}^n$, its corresponding quantum state $| x \rangle$ is defined as $\frac{1}{\|x\|} \sum_{i = 1}^n x_i |i - 1\rangle$, where the computational basis state $|i - 1\rangle$ is associated with a probability proportional to $x_i^2$. However, without considering this quantum state motivation, the most natural way to assign probabilities to data elements is to make them proportional to the absolute values of the elements, rather than their squares. For a systematic comparison, we first extend Section~\ref{sec:background} to arbitrary norms.

\subsection{$\text{SQ}_p$ data structure}
Figure~\ref{fig:lp_diagram} with arbitrary $p \geq 1$ is a direct generalization of $\text{SQ}_2$, which we refer to as $\text{SQ}_p$.

\begin{definition}
\label{def:lp_distribution}
For a nonzero vector $x \in \mathbb{R}^n$, we denote by $\mathcal{D}^{(p)}_x$ the distribution over $[n]$ with PDF

\begin{equation*}
\mathcal{D}^{(p)}_x(i) = \frac{|x_i|^p}{||x||_p^p}.
\end{equation*}
\end{definition}

\noindent Hereafter, the notation $\mathcal{D}^{(p)}_x$ implies that $x$ is a nonzero vector. The generalizations of Lemma~\ref{lemma:vector_structure_exists} and Corollary~\ref{corollary:matrix_structure_exists} follow simply by replacing $\mathcal{D}_x$, $\mathcal{D}_{\Tilde{A}}$, and $\mathcal{D}_{A^{(j)}}$ with $\mathcal{D}^{(p)}_x$, $\mathcal{D}^{(p)}_{\Tilde{A}^{(p)}}$, and $\mathcal{D}^{(p)}_{A^{(j)}}$, respectively. While the analysis under this generalization is relatively straightforward based on prior studies, we elaborate on it in the following sections to express complexities in terms of $p$.

\subsection{Inner product estimation with $\text{SQ}_p$}
\label{sec:lp_inner_product}

For $v \in \mathbb{R}^n$, define

\begin{equation*}
v^{(p)} := (|v_i|^p)_{i = 1}^{n} \in \mathbb{R}^n.
\end{equation*}
We state a generalization of Proposition~\ref{prop:l2_inner_product_prop} to the $L^p$ data structure.

\begin{observation}
\label{prop:general_inner_product}
Given $\textnormal{SQ}_p(x)$ and $\textnormal{Q}(y)$ for $x, y \in \mathbb{R}^n$, the inner product $\langle x, y \rangle$ can be estimated to additive error $\epsilon ||x||_p^{p / 2} \sqrt{\langle x^{(2 - p)}, y^{(2)} \rangle}$ with at least $1 - \delta$ probability using $O\left(\frac{1}{\epsilon^2} 
\log \frac{1}{\delta} \right)$ queries and samples.
\end{observation}

\noindent\textit{Proof.} See Appendix~\ref{sec:proof_of_proposition_general_inner_product}.

As a practical example, the quantum-inspired recommendation algorithm involves estimating the inner products between user rows \citep{tang2019quantum}. Observation~\ref{prop:general_inner_product} was evaluated on the MovieLens dataset \citep{harper2015movielens}, comprising 330,975 users and their ratings for 288,983 movies. The sampled pairs of rows were restricted to those with a minimum of 50 shared nonzero positions. For each pair, the constant $||x||_p^{p / 2} \sqrt{\langle x^{(2 - p)}, y^{(2)} \rangle}$ in Observation~\ref{prop:general_inner_product}, which is directly related to the number of samples required for a fixed precision, was evaluated for $p = 1, 2$. On average, this constant was $\approx 2.16$ times smaller for $p = 1$ compared to $p = 2$, which implies that $\text{SQ}_1$ requires fewer samples than $\text{SQ}_2$ to achieve the same accuracy in estimating the inner product.

Assuming certain conditions, theoretical support can be provided for the experimental evidence showing that $\text{SQ}_1$ is preferable to $\text{SQ}_2$. Note that in Observation~\ref{prop:general_inner_product}, while $x$ is fixed for most of the time, the input $y$ varies across different problem instances. Therefore, it makes sense to analyze the average performance by viewing $y$ as a random variable. If we make further assumptions about the input $y$, the performance (averaged over $y$) can be expressed only in terms of $x$ and $p$, which determines the optimal value of $p$. In this respect, we now focus on deriving the optimal sample access under specific, simplified conditions.

In machine learning, processing steps like normalization are frequently applied to input data \citep{ioffe2015batch}. Therefore, we assume that the entries of the input vector have equal second-order moments. Even without this assumption, the law of large numbers and concentration phenomena suggest that deviations in the algorithm's global behavior are likely negligible, since we operate on high-dimensional vectors.

\begin{observation}
\label{proposition:inner_product_estimation}
$\textnormal{SQ}_1$ has the optimal sample efficiency, averaged over $y$, for inner product estimation in Observation~\ref{prop:general_inner_product} if the entries of $y$ have equal second-order moments.
\end{observation}

\noindent\textit{Proof.} See Appendix~\ref{sec:proof_of_proposition_inner_product_estimation}.

\begin{observation}
\label{corollary:gaussian_improvement}
If the elements of $x$ and $y$ are i.i.d. zero mean Gaussian (uniform) random variables, replacing $\textnormal{SQ}_2$ with $\textnormal{SQ}_1$ improves the sample efficiency by a factor of $\pi / 2$ ($4 / 3$) on average.
\end{observation}
\noindent\textit{Proof.} See Appendix~\ref{appendix:improvement}.

\begin{remark}
In addition to sample efficiency, it can be seen from Appendix~\ref{sec:proof_of_proposition_general_inner_product} that $\textnormal{SQ}_1$ has a further computational advantage over $\textnormal{SQ}_2$ for inner product estimation because it does not require floating-point division in calculating each sample.
\end{remark}
\begin{remark}
\label{remark:trace_inner_product_estimation}
If we are given $\textnormal{SQ}_p(A)$ for some matrix $A$, then estimating the value of $x^{\top} A y$ for \textit{two} inputs $x$ and $y$ with query access is also an inner product estimation. Indeed, this is a special case of \textit{trace inner product estimation} introduced in the context of the quantum-inspired support vector machine \citep{ding2021quantum}. Under the assumption that each $x_i \cdot y_j$ has the same second-order moment, Observation~\ref{proposition:inner_product_estimation} holds for this task.
\end{remark}

\subsection{Sampling from a linear combination of vectors with $\text{SQ}_p$}

The sampling method that achieves the result of Proposition~\ref{prop:l2_linear_combination} utilizes \textit{rejection sampling} \citep{forsythe1972neumann, neal2003slice, casella2004generalized, thomopoulos2012essentials, legault2019accounting}. Suppose $P(i)$ is the distribution to which we have sample access and $Q(i)$ is the target distribution. Each time we pull a sample $i$ from $P$, we compute the ratio $r_i := \frac{Q(i)}{M P(i)}$ for some constant $M$. Then we output $i$ with probability $r_i$ and restart (or reject) otherwise.

We generalize Proposition~\ref{prop:l2_linear_combination} to the $L^p$ data structure.

\begin{observation}
\label{prop:general_linear_combination}
Suppose $p \geq 1$. Given $\textnormal{SQ}_p(A)$ for $A \in \mathbb{R}^{m \times n}$ and $\textnormal{Q}(x)$ for $x \in \mathbb{R}^n$, we can output a sample from $\mathcal{D}_{Ax}^{(p)}$ in $O(n^p C^{(p)}(A, x))$ expected query complexity, where

\begin{equation*}
C^{(p)}(A, x) := \frac{\sum_{j = 1}^{n} ||x_j A^{(j)}||_p^p}{||Ax||_p^p}.
\end{equation*}
\end{observation}

\noindent\textit{Proof.} See Appendix~\ref{sec:proof_of_proposition_general_linear_combination}.

With a slight abuse of notation, define
\begin{equation*}
M(p) := n^{p - 1} \frac{\sum_{j = 1}^{n} ||x_{j} A^{(j)}||_p^p}{||Ax||_p^p}.
\end{equation*}
The expected query complexity is then $O(n M(p))$.

\begin{observation}
\label{corollary:sparse_no_cancellation}
If the columns of $A$ are sparse such that their supports have negligible overlap, then $C^{(p)}(A, x) \approx 1$ and $M(p) \approx n^{p - 1}$.
\end{observation}

In this section, we analyze performance purely in terms of expected number of queries per sample. Other factors are discussed in Section~\ref{section:discussion}. We consider the recommendation system again. Each item recommended to a user is sampled from a linear combination of a small subset of relevant user rows in the MovieLens dataset. The number of selected users ranges from 10 to 100, and the coefficients for the linear combination were drawn independently from $\mathcal{N}(0, 1)$. As observed in Table~\ref{tab:lin_comb_movielens}, there is a noticeable gap in the expected number of queries per sample between $\text{SQ}_2$ and $\text{SQ}_1$. This is because the user rows are sparse and $M(p)$ behaves like $n^{p - 1}$ (see Observation~\ref{corollary:sparse_no_cancellation}).

\begin{table}[h]
\caption{The ratio of the expected number of queries per sample when using $\text{SQ}_2$ to that when using $\text{SQ}_1$ (which is equal to $M(2) / M(1)$) for the MovieLens dataset, averaged over 1000 instances. As in Observation~\ref{prop:general_linear_combination}, $n$ denotes the dimension of $x$, or equivalently, the number of users involved in the linear combination of user vectors in the context of recommendation systems.}
\label{tab:lin_comb_movielens}
\begin{tabular}{c|c|c|c|c|c|c|c|c|c|c}
\toprule
$n$ & 10 & 20 & 30 & 40 & 50 & 60 & 70 & 80 & 90 & 100 \\
\midrule
Ratio & 8.99 & 16.0 & 23.0 & 28.0 & 32.4 & 37.0 & 43.2 & 46.3 & 48.0 & 52.5 \\
\botrule
\end{tabular}
\end{table}

As in Section~\ref{sec:lp_inner_product}, we provide an analysis supporting the preference for $\text{SQ}_1$ when the sparsity assumption is relaxed. Recall that in Section~\ref{sec:lp_inner_product}, it was shown both experimentally and mathematically that $\text{SQ}_1$ is likely to achieve better sample efficiency than $\text{SQ}_2$. We hope this advantage extends to the current task, i.e., $M(p) \geq M(1)$. Under a very limited condition, this is always true. Let us illustrate a warm-up example.

\begin{observation}
\label{lemma:positive}
For $x \in \mathbb{R}^n$, define
\begin{equation*}
\textnormal{sgn}(x) := (\textnormal{sgn}(x_i))_{i = 1}^{n} \in \mathbb{R}^n
\end{equation*}
and assume that $\textnormal{sgn}(A_i) \in \{ \textnormal{sgn}(x), -\textnormal{sgn}(x) \}$ for all $i \in [m]$, where $A_i$ is the $i$-th row of $A$. Then for $p \geq 1$ we have $M(p) \geq M(1)$.
\end{observation}

\noindent\textit{Proof.} See Appendix~\ref{sec:proof_of_observation_positive}.

A counterexample also exists. That is, we can find $A$ and $x$ such that $M(p) < M(1)$ for some $p > 1$. For example, let
\begin{equation*}
A = \begin{pmatrix}
1 & 1\\
2 & -2
\end{pmatrix}, \quad x = \begin{pmatrix}
1\\
-1
\end{pmatrix}.
\end{equation*}
A simple calculation gives $M(1) = 3 / 2 > M(2) = 5 / 4$.

Now, we claim that under certain assumptions, $\text{SQ}_1$ is optimal on average for sampling from a linear combination of vectors. The average is taken over both $A$ and $x$. To simplify the mathematical analysis, from here on, we assume that the entries of $A$ and $x$ are i.i.d. zero mean random variables following a PDF $f$ and that $m, n \gg 1$. Note that it is more reasonable to assume that the entries of $A$ follow a distribution distinct from that of $x$. Moreover, if $A$ reflects a type of nonnegative quantities, such as rankings in recommendation systems, its entries will have a nonzero mean. Due to the difficulty of deriving a neat formula for such cases, they will be addressed experimentally in Section~\ref{sec:experimental_results}. Define
\begin{gather*}
\sigma_f^2 := \text{Var}_{X \sim f}[X], \quad \mu_{f, p} := \mathbb{E}_{X \sim f}[|X|^p], \quad \sigma_{f, p}^2 := \text{Var}_{X \sim f}[|X|^p],\\
\Tilde{\mu}_{f, p} := \mathbb{E}_{X \sim \mathcal{N}(0, \sigma_f^4)}[|X|^p] = \sigma_f^{2p} 2^{p / 2} \frac{\Gamma(\frac{p + 1}{2})}{\sqrt{\pi}}.
\end{gather*}
We want to express $\mathbb{E} [M(p) / n^{p / 2}]$ as a function of $p$ as $m, n \rightarrow \infty$. Formally, we state the following lemma.

\begin{observation}
\label{lemma:rigorous_proof}
If the entries of $A$ and $x$ are i.i.d. zero mean random variables following a distribution $f$, then $M(p) / n^{p / 2}$ converges in probability to $\mu_{f, p}^2 / \Tilde{\mu}_{f, p}$ as $m, n \rightarrow \infty$.
\end{observation}

\noindent\textit{Proof.} See Appendix~\ref{sec:proof_of_lemma_rigorous_proof}.

Suppose we fix the value of $p$. Then from Observation~\ref{lemma:rigorous_proof}, the expected number of iterations for a single valid sample, i.e., $M(p)$, behaves like $n^{p / 2}$ times a constant for large $n$. Therefore, the expected query complexity for a single valid sample is $O(n^{1 + p / 2})$. It is evident from this expression that small $p$ is desirable. The following proposition immediately follows.

\begin{observation}
\label{proposition:sampling_from_linear_comb}
If the entries of $A$ and $x$ are i.i.d. zero mean random variables and $m, n \gg 1$, then $\textnormal{SQ}_1$ minimizes on average the expected number of queries per sample for sampling from a linear combination of vectors in Observation~\ref{prop:general_linear_combination}, where the average is taken over $A$ and $x$.
\end{observation}

In particular, the expected number of iterations for a single valid sample for $\textnormal{SQ}_1$ is $O(1 / \sqrt{n})$ times that for $\textnormal{SQ}_2$ (i.e., $O(\sqrt{n})$ times faster).

Observations~\ref{proposition:inner_product_estimation} and \ref{proposition:sampling_from_linear_comb} collectively suggest that for the tasks of inner product estimation and sampling from a linear combination of vectors, $\textnormal{SQ}_1(x)$ is preferable to $\text{SQ}_2$ in terms of time complexities. However, it is important to note that Observation~\ref{proposition:sampling_from_linear_comb} compares $\textnormal{SQ}_1$ with other data structures solely in terms of time complexity. If the underlying norm itself used for sampling matters, a fair comparison cannot be made. This issue is discussed further in Section~\ref{section:discussion}.

\section{Numerical simulations}
\label{sec:experimental_results}
Observation~\ref{lemma:rigorous_proof} assumes $m \gg 1$ and $n \gg 1$. While the former is a reasonable assumption for quantum-inspired algorithms to have practical advantages, the latter might not be the case (consider the recommendation scenario). We plot the empirical mean of $M(p)$ for small values of $n$. The following two distributions are considered for $f$ in Observation~\ref{lemma:rigorous_proof}:
\begin{itemize}
\item $f = \mathcal{N}(0, 1)$, where $\sigma_f^2 = 1$, $\mu_{f, p} = \Tilde{\mu}_{f, p} = 2^{p / 2} \Gamma\left(\frac{p + 1}{2}\right) / \sqrt{\pi}$.
\item $f = \mathcal{U}_{(-1, 1)}$, where $\sigma_f^2 = 1 / 3$, $\mu_{f, p} = 1 / (p + 1)$, $\Tilde{\mu}_{f, p} = (\sqrt{2} / 3)^p \Gamma\left(\frac{p + 1}{2}\right) / \sqrt{\pi}$.
\end{itemize}

Figures~\ref{fig:many_figures1} \& \ref{fig:many_figures2} (see Appendix~\ref{section:many_figures_appendix}) show the results for $f = \mathcal{N}(0, 1)$ and $f = \mathcal{U}_{(-1, 1)}$, respectively. As expected, for $n \gg 1$, the estimated value of $\mathbb{E} [M(p)]$ (i.e., assuming $n \rightarrow \infty$) closely matches its true (i.e., empirical) value. Even for small $n$, the numerical results suggest that $\mathbb{E}[M(p)]$ is an increasing function of $p$.

\begin{table}[h]
\caption{The ratio of the expected number of queries per sample when using $\text{SQ}_2$ to that when using $\text{SQ}_1$ (which is equal to $M(2) / M(1)$) averaged over 1000 iterations across different distributions of the elements of $A$, with $m$ is fixed at 1024.}\label{tab:distributions_m2m1}
\footnotesize
\setlength{\tabcolsep}{4.5pt}
\begin{tabular}{l|c|c|c|c|c|c|c|c|c|c|c|c}
\toprule
\textbf{Distribution} \textbackslash \, $n$ & 2 & 4 & 8 & 16 & 32 & 64 & 128 & 256 & 512 & 1024 & 2048 & 4096 \\
\midrule
$\mathcal{N}(0, 1)$ & 1.58 & 2.39 & 3.46 & 4.97 & 7.04 & 9.99 & 14.2 & 20.1 & 28.3 & 40.1 & 56.8 & 80.3 \\
$\mathcal{U}_{(-1, 1)} \vphantom{a}$ & 1.55 & 2.26 & 3.26 & 4.61 & 6.53 & 9.23 & 13.1 & 18.5 & 26.2 & 36.9 & 52.3 & 74.0 \\
$\text{Laplace}(0, 1)$ & 1.63 & 2.53 & 3.77 & 5.48 & 7.87 & 11.2 & 16.0 & 22.6 & 32.0 & 45.4 & 64.1 & 90.6 \\
$\text{Laplace}(1, 1)$ & 1.60 & 2.46 & 3.62 & 5.25 & 7.55 & 10.7 & 15.2 & 21.6 & 30.7 & 43.3 & 60.7 & 86.0 \\
$\text{Exp}(1)$ & 1.70 & 2.76 & 4.19 & 6.19 & 8.87 & 12.6 & 18.2 & 25.7 & 36.4 & 50.9 & 72.9 & 103 \\
$\text{Beta}(2, 2)$ & 1.83 & 3.04 & 4.42 & 6.47 & 9.27 & 13.4 & 18.6 & 25.8 & 38.1 & 53.5 & 76.5 & 108 \\
$\text{Beta}(3, 3)$ & 1.92 & 3.12 & 4.77 & 6.81 & 9.77 & 13.9 & 19.6 & 27.3 & 39.4 & 55.1 & 80.6 & 112 \\
$\text{Beta}(2, 1)$ & 1.95 & 3.15 & 4.76 & 6.86 & 10.2 & 14.1 & 19.9 & 28.7 & 40.8 & 56.5 & 80.2 & 113 \\
$\text{Beta}(1, 2)$ & 1.74 & 2.81 & 4.25 & 6.11 & 8.79 & 12.3 & 17.6 & 24.8 & 34.4 & 50.3 & 70.1 & 99.2 \\
$\text{Beta}(5, 2)$ & 2.16 & 3.64 & 5.47 & 8.14 & 11.6 & 16.8 & 23.5 & 33.5 & 47.4 & 67.5 & 96.6 & 133 \\
$\text{Beta}(2, 5)$ & 1.77 & 2.87 & 4.28 & 6.17 & 8.94 & 12.5 & 18.1 & 25.1 & 35.9 & 51.5 & 70.8 & 101 \\
$\text{Gamma}(2, 2)$ & 1.70 & 2.71 & 4.07 & 5.99 & 8.61 & 12.3 & 17.5 & 24.6 & 35.4 & 49.5 & 70.5 & 100 \\
$\text{Gamma}(0.5, 1)$ & 1.76 & 2.90 & 4.59 & 6.86 & 9.94 & 14.5 & 20.7 & 29.2 & 41.6 & 58.8 & 83.7 & 118 \\
$\text{Gamma}(10, 0.5)$ & 2.00 & 3.27 & 4.89 & 7.09 & 10.3 & 14.4 & 20.9 & 29.4 & 42.6 & 60.0 & 83.7 & 118 \\
\botrule
\end{tabular}
\end{table}

Next, the runtimes of $\text{SQ}_2$ and $\text{SQ}_1$ are evaluated across various distributions of the elements of $A$, including PDFs with nonzero mean. The coefficients for the linear combination were drawn independently from $\mathcal{N}(0, 1)$. From Table~\ref{tab:distributions_m2m1}, it can be seen that $\text{SQ}_1$ is consistently faster than $\text{SQ}_2$.

\section{Applications to DFE}
\label{sec:dfe_application}

As another example, we explain how sample access affects DFE using Pauli measurements \citep{flammia2011direct, huang2020predicting, zhang2021direct, leone2023nonstabilizerness}. DFE enables efficient and scalable verification of quantum states in quantum computing devices by circumventing the exponential resource requirements of full quantum state tomography \citep{da2011practical, flammia2011direct}. It is essentially a sampling-based inner product estimation algorithm. In this sense, it aligns with the principles of quantum-inspired machine learning. This prompts us to investigate whether sampling based on $L^1$ norm improves the performance of DFE. We provide an overview of the prerequisites and frameworks.

Suppose our goal is to prepare an $n$-qubit quantum state (with Hilbert space dimension $d = 2^n$) in the lab to be a \text{pure} target state $\rho = |\psi\rangle\langle\psi|$, but we have instead produced an \textit{unknown} state $\sigma$. It is typical to use \textit{fidelity} as a metric to evaluate how similar $\sigma$ is to $\rho$ \citep{jozsa1994fidelity, nielsen2010quantum, wilde2013quantum, liang2019quantum, baldwin2023efficiently}. If the target state $\rho$ is pure, the fidelity between $\rho$ and $\sigma$ is defined as

\begin{equation*}
F(\rho, \sigma) \equiv \text{tr}(\rho \sigma).
\end{equation*}
For each $k \in \{1, 2, \cdots , d^2\}$, define the characteristic function $\chi_{\rho}(k) \equiv \text{tr}(\rho W_k / \sqrt{d})$, where $W_k \in \{\mathbb{I}, X, Y, Z\}^{\otimes n}$ is an $n$-qubit Pauli operator. Then

\begin{equation}
\label{equation:characteristic_formulation}
\text{tr}(\rho \sigma) = \sum_k \chi_{\rho}(k) \chi_{\sigma}(k).
\end{equation}
Note that (\ref{equation:characteristic_formulation}) is in the form of a vector inner product. Clearly, calculating each summand in (\ref{equation:characteristic_formulation}) is impractical since $d^2$ scales exponentially with $n$. For this reason, the exact same technique as in Proposition~\ref{prop:l2_inner_product_prop} was independently developed for the task of estimating (\ref{equation:characteristic_formulation}). We select $k \in \{1, 2, \cdots , d^2\}$ with probability $(\chi_{\rho}(k))^2$ and define the estimator as $X \equiv \chi_{\sigma}(k) / \chi_{\rho}(k)$. Normalization is not necessary since $\sum_k (\chi_{\rho}(k))^2 = 1$. The difference from Proposition~\ref{prop:l2_inner_product_prop} is that we can only estimate $\chi_{\sigma}(k)$ rather than having query access to it.

\subsection{Sample efficiency for the W state}
\label{section: w_state_section}
In this section, we will restrict our focus to the W state \citep{dur2000three, cabello2002bell} as the target:

\begin{equation*}
\rho = \frac{1}{n}\left(\sum_{i = 1}^n |e_i\rangle\langle e_i| + \sum_{i \neq j} |e_i\rangle\langle e_j|\right),
\end{equation*}
where
\begin{equation*}
|e_i\rangle \equiv |\underbrace{00 \cdots 0}_{\times (i - 1)} 1 \underbrace{00 \cdots 0}_{\times (n - i)}\rangle
\end{equation*}
and $n \geq 3$. It is known that from

\begin{equation}
\label{equation:original_bound}
N \leq \frac{2\log (2 / \delta)}{\epsilon^2}n^2 + \frac{1}{\epsilon^2 \delta} + 1
\end{equation}
random Pauli measurements, DFE can achieve an additive error of $2\epsilon$ with a success probability of at least $1 - 2\delta$, where we \textit{fix} the parameters $(\epsilon, \delta)$.

Now suppose we sample $k$ from $\text{SQ}_1$, instead of $\text{SQ}_2$. While the assumption in Observation~\ref{proposition:inner_product_estimation} does not apply to the current task, it turns out that we can independently show that $\text{SQ}_1$ tightens the upper bound on the number of samples needed to achieve the same precision. First, select $k$ with probability $|\chi_{\rho}(k)| / Z$, where $Z \equiv \sum_k |\chi_{\rho}(k)|$ is the normalizing constant. Then define the estimator as
\begin{equation*}
\label{equation:X_definition}
X \equiv Z \text{sgn}(\chi_{\rho}(k))\chi_{\sigma}(k).
\end{equation*}
Each Pauli operator $W$ can be represented by two binary vectors $\textbf{j}, \textbf{k} \in \{0, 1\}^n$ which denote the positions for $X$ and $Z$, respectively (up to a global phase). For example, $W = \mathbb{I} \otimes X \otimes Y \otimes Z$ corresponds to $\textbf{j} = (0, 1, 1, 0)$ and $\textbf{k} = (0, 0, 1, 1)$ (note that $Y = iXZ$). Sampling $k$ from $\text{SQ}_1$ is equivalent to sampling $\textbf{j}, \textbf{k}$ from the following distribution, which directly follows from \citet[Equation (27)]{flammia2011direct}:
\begin{equation}
\label{equation:l1_probabilities}
p(\textbf{j}, \textbf{k}) =
\begin{cases} 
\frac{1}{n\sqrt{d}} |n - 2|\textbf{k}|| & \text{if } \textbf{j} = \bm{0}, \\
\frac{2}{n\sqrt{d}} & \text{if } |\textbf{j}| = 2 \text{ and } \textbf{j} \cdot \textbf{k} \bmod 2 = 0 \\
0 & \text{otherwise.}
\end{cases}
\end{equation}
For $l = \left\lceil 1 / \epsilon^2 \delta \right\rceil$, let
\begin{equation*}
Y = \frac{1}{l}\sum_{i = 1}^l X_i = \frac{1}{l}\sum_{i = 1}^l Z \text{sgn}(\chi_{\rho}(k_i))\chi_{\sigma}(k_i),
\end{equation*}
where each $k_i$ is sampled according to (\ref{equation:l1_probabilities}). Also, for each $1 \leq i \leq l$, we perform

\begin{equation*}
N_i = \left\lceil \frac{1}{2l\epsilon^2} \log (2 / \delta) n^2 \right\rceil
\end{equation*}
independent measurements and use the outcomes $\{M_{ij}\}_{j = 1}^{N_i}$ to estimate $X_i$:
\begin{equation*}
\label{equation:tilde_X_definition}
\Tilde{X}_i = \frac{Z \text{sgn}(\chi_{\rho}(k_i))}{\sqrt{d}} \frac{1}{N_i} \sum_{j = 1}^{N_i} M_{ij}.
\end{equation*}
Then we can estimate $Y$ as

\begin{equation}
\label{equation:final_estimator}
\Tilde{Y} = \frac{1}{l} \sum_{i = 1}^l \Tilde{X}_i = \sum_{i = 1}^l \sum_{j = 1}^{N_i} \frac{Z \text{sgn}(\chi_{\rho}(k_i))}{l N_i \sqrt{d}}M_{ij},
\end{equation}
where $\mathbb{E}[\Tilde{Y}] = Y$. Since each $M_{ij}$ is $\pm 1$, each sample in (\ref{equation:final_estimator}) has an absolute value of $Z / l N_i \sqrt{d}$. We can bound this value by upper-bounding $Z$ (see Appendix~\ref{appendix:derive_Z_upper_bound}):

\begin{align}
\label{equation:Z_upper_bound}
Z \leq \left(\frac{n}{2} + \frac{1}{\sqrt{n}} - \frac{1}{2}\right) \sqrt{d}.
\end{align}
Meanwhile, a direct calculation gives $Z = 3\sqrt{2}$ for $n = 3$, so we conclude that $Z \leq (n / 2)\sqrt{d}$ for $n \geq 3$. Therefore, we have $Z / l N_i \sqrt{d} \leq n / 2 l N_i$.

As in \cite{flammia2011direct}, one can show that
\begin{equation}
\label{equation:l_level_estimation}
\text{Pr}[|Y - \text{tr}(\rho\sigma)| \geq \epsilon] \leq \delta
\end{equation}
and

\begin{equation}
\text{Pr}[|\Tilde{Y} - Y| \geq \epsilon] \leq \delta,
\end{equation}
which gives the desired accuracy. The total number of required measurements is

\begin{equation}
\label{equation:total_number_of_samples}
N = \sum_{i = 1}^l N_i = l \left\lceil \frac{1}{2l\epsilon^2} \log (2 / \delta) n^2 \right\rceil \leq \frac{\log (2 / \delta)}{2\epsilon^2}n^2 + l \leq \frac{\log (2 / \delta)}{2\epsilon^2}n^2 + \frac{1}{\epsilon^2 \delta} + 1,
\end{equation}
which improves the bound (\ref{equation:original_bound}) asymptotically by a factor of $4$.

To sample from $\text{SQ}_1$ for the characteristic function $\chi_{\rho} (k)$, we must calculate the exact value of $Z$. From the binomial identity
\begin{equation}
\label{equation:binomial_identity}
Z' \equiv \sum_{w = 0}^n \binom{n}{w} |n - 2w| = 2n\binom{n - 1}{\lfloor n / 2 \rfloor}
\end{equation}
(see Appendix~\ref{appendix:binomial}), we have
\begin{equation*}
Z = \frac{2}{\sqrt{d}}\binom{n - 1}{\lfloor n / 2 \rfloor} + \frac{(n - 1)\sqrt{d}}{2}.
\end{equation*}
Each $k_i$ is sampled as follows. With probability $(n - 1)\sqrt{d} / 2Z$, go to the second branch in (\ref{equation:l1_probabilities}); otherwise, go to the first branch. If we are in the first branch, sample $w$ with probability $\binom{n}{w}|n - 2w| / Z'$. The remaining steps are the same as in \cite{flammia2011direct}.
\begin{remark}
For stabilizer states (e.g., the GHZ state \citep{greenberger1989going, mermin1990quantum, caves2002unknown}), the characteristic function yields a uniform distribution over $d$ operators. In this case, replacing $\textnormal{SQ}_2$ with $\textnormal{SQ}_1$ makes no difference.
\end{remark}
\subsection{Sample efficiency for general well-conditioned states}
We derive a more general characterization of $\text{SQ}_1$ sampling for the family of \textit{well-conditioned} states.
\begin{definition}[Well-conditioned state \citep{flammia2011direct}]
A state $\rho$ is well-conditioned with parameter $\alpha$ if for all $k$, either $\text{tr}(\rho W_k) = 0$ or $|\text{tr}(\rho W_k)| \geq \alpha$.
\end{definition}
Stabilizer states (e.g., the GHZ state), the W state, and the Dicke state with $k$ excitations \citep{kiesel2007experimental, bartschi2019deterministic} are well-conditioned with $\alpha = 1$, $\alpha = 1 / n$, and $\alpha = O(1 / n^k)$, respectively.

Note that (\ref{equation:l_level_estimation}) holds for either $l = \lceil 1 / \epsilon^2 \delta \rceil$ or $l = \lceil 2 \log (2 / \delta) / \alpha^2 \epsilon^2 \rceil$. If $(\epsilon, \delta)$ is fixed and $1 / \alpha$ scales with $n$, the former is preferred for large $n$, as the value of $l$ itself contributes to the total number of samples (\ref{equation:total_number_of_samples}). We show that in this case, $\text{SQ}_1$ sampling is at least as good as $\text{SQ}_2$ sampling. Formally, we establish the following statement.

\begin{proposition}
\label{proposition:class_of_states}
Let $\{\rho (n)\}_n$ denote a family of quantum states, defined for each $n$. If $\rho (n)$ is well-conditioned with parameter $\alpha (n)$ and $1 / \alpha (n) = \omega (1)$, then the upper bound on the total number of Pauli measurements for $\textnormal{SQ}_1$  is asymptotically (with respect to $n$) less than or equal to that for $\textnormal{SQ}_2$.
\end{proposition}

\noindent\textit{Proof.} See Appendix~\ref{sec:proof_of_proposition_class_of_states}.

\section{Discussion}
\label{section:discussion}
For applications where inner product estimation and/or sampling from a linear combination of vectors dominate the time complexity, it is likely that $\text{SQ}_1$ is the best choice. However, the general implication of our observations is not that $\text{SQ}_1$ is universally optimal. Rather, they indicate that we should carefully determine the optimal value of $p$ in $\text{SQ}_p$ each time we develop a new quantum-inspired algorithm. The first and most obvious reason is the possibility that the overall time complexity may be dominated by factors beyond the optimization of the subroutines considered in this work. Secondly, while this issue might not arise with inner product estimation, we must consider if it is reasonable to generate a sample based on $p$-norm when sampling from a linear combination of vectors. If the sample generation itself serves as the final output of an algorithm (rather than a computational tool), and there is a compelling reason to prefer $L^2$ norm sampling over $L^1$ norm sampling, trade-offs between output quality and time complexity may arise. Nevertheless, we argue that this trade-off is worth considering, as the sample efficiency using $\text{SQ}_1$ is asymptotically better in terms of the number of vectors involved in the linear combination.

\section{Conclusion}
\label{sec:conclusion}
Quantum-inspired algorithms achieve polylogarithmic time complexities with the help of a BST-like data structure. We have shown that storing the data based on $L^1$ norm is preferred. This result follows from analyzing two important subroutines, which constitute computational bottlenecks in many quantum-inspired algorithms, by generalizing $\text{SQ}_2$ data structures to $\text{SQ}_p$ data structures.

A quantity in the form of an inner product can be efficiently estimated using importance sampling, even for vectors of extremely large dimensions. We can also quickly draw a sample from a linear combination of high-dimensional vectors. Notable applications include recommendation systems and DFE, which involve high-dimensional vector operations or exhibit exponential growth in vector dimension with respect to the problem size. For these tasks, we observe that sampling vector indices based on $L^1$ norm results in improved sample efficiency. Investigating the optimal data structures for existing or upcoming quantum and quantum-inspired algorithms presents a compelling avenue for future research.

\begin{appendices}
\renewcommand{\thefigure}{\arabic{figure}}
\setcounter{figure}{2}

\section{Proof of Observation~\ref{prop:general_inner_product}}
\label{sec:proof_of_proposition_general_inner_product}

\begin{proof}
We use the \textit{importance sampling} technique \citep{goertzel1949quota, kahn1951estimation, kloek1978bayesian}. Define a random variable

\begin{equation*}
Z(i) := \text{sgn}(x_i) |x_i|^{1 - p} y_i,
\end{equation*}
where $i \sim \mathcal{D}^{(p)}_x$. Notice that $Z(i)$ will never be sampled if $x_i = 0$, so we need not be concerned about negative powers of zero in the expression $|x_i|^{1 - p}$. Then we have
\begin{align*}
\mathbb{E}[Z(i)]_{i \sim \mathcal{D}^{(p)}_x} & = \sum_{i = 1}^{n} \frac{|x_i|^p}{||x||_p^p} \text{sgn}(x_i) |x_i|^{1 - p} y_i\\
& = \frac{\sum_{i = 1}^{n} x_i y_i}{||x||_p^p}\\
& = \frac{\langle x, y \rangle}{||x||_p^p}
\end{align*}
and

\begin{align*}
\text{Var}[Z(i)]_{i \sim \mathcal{D}^{(p)}_x} & \leq \mathbb{E}[Z(i)^2]_{i \sim \mathcal{D}^{(p)}_x}\\
& = \sum_{i = 1}^{n} \frac{|x_i|^p}{||x||_p^p} |x_i|^{2 - 2p} y_i^2\\
& = \frac{\sum_{i = 1}^{n} |x_i|^{2 - p} y_i^2}{||x||_p^p}\\
& = \frac{\langle x^{(2 - p)}, y^{(2)} \rangle}{||x||_p^p}.
\end{align*}
Therefore, the random variable $||x||_p^p Z(i)$ has mean $\langle x, y \rangle$ and standard deviation less than or equal to $||x||_p^{p / 2} \sqrt{\langle x^{(2 - p)}, y^{(2)} \rangle}$. It is known that the median of $6 \log \frac{1}{\delta}$ sample means with each sample containing $\frac{9}{2 \epsilon^2}$ data points is $\epsilon \sigma$-close to $\mu$ with probability at least $1 - \delta$, where $\mu$ and $\sigma$ are the mean and variance of the random variable, respectively. The result directly follows.
\end{proof}

\section{Proof of Observation~\ref{proposition:inner_product_estimation}}
\label{sec:proof_of_proposition_inner_product_estimation}

We need the following lemma.
\begin{lemma}
\label{lemma:attain_minimum}
For $x \in \mathbb{R}^n$ and $p \in \mathbb{R}$, the function

\begin{equation}
\label{equation:f_p_definition}
f(p) := ||x||_p^p ||x||_{2 - p}^{2 - p}
\end{equation}
attains its minimum at $p = 1$.
\end{lemma}

\begin{proof}
The statement is trivially true if $x$ is a zero vector. Now assume that $x$ is a nonzero vector. Without loss of generality, we further assume that each entry of $x$ is nonzero, since if $x_i = 0$ then it has no contribution to (\ref{equation:f_p_definition}). For $p \in \mathbb{R}$, define

\begin{equation*}
g(p) := \frac{\sum_{i = 1}^{n} |x_i|^p \ln{|x_i|}}{\sum_{i = 1}^{n} |x_i|^p}.
\end{equation*}
We have

\begin{equation*}
g'(p) := \frac{\sum_{i = 1}^{n} |x_i|^p (\ln{|x_i|})^2 \sum_{i' = 1}^{n} |x_{i'}|^p - \left( \sum_{i = 1}^{n} |x_i|^p \ln{|x_i|} \right)^2}{\left(\sum_{i = 1}^{n} |x_i|^p\right)^2}.
\end{equation*}
From Cauchy-Schwarz inequality,

\begin{align*}
\sum_{i = 1}^{n} |x_i|^p (\ln{|x_i|})^2 \sum_{i' = 1}^{n} |x_{i'}|^p & \geq \left( \sum_{i = 1}^{n} |x_i|^{p / 2} (\ln{|x_i|}) |x_i|^{p / 2} \right)^2\\
& = \left( \sum_{i = 1}^{n} |x_i|^p \ln{|x_i|} \right)^2,
\end{align*}
where equality holds for all $p$ if and only if $\forall i, \, x_i = \pm e^{\lambda}$ for some $\lambda \in \mathbb{R}$ and does not hold for all $p$ otherwise. In the former case, $f(p) = n e^{p\lambda} n e^{(2 - p)\lambda} = n^2 e^{2\lambda}$ is constant. In the latter case, we have $g'(p) > 0$ for all $p$. Now we differentiate $f$:

\begin{equation}
\label{equation:derivative}
f'(p) = \sum_{i = 1}^{n} |x_i|^p \ln{|x_i|} \sum_{i' = 1}^{n} |x_{i'}|^{2 - p} - \sum_{i = 1}^{n} |x_i|^p \sum_{i' = 1}^{n} |x_{i'}|^{2 - p} \ln{|x_{i'}|}.
\end{equation}
Assume that $f$ attains its minimum at $\hat{p} \neq 1$. Then it must be that $f'(\hat{p}) = 0$. From (\ref{equation:derivative}), we have

\begin{equation*}
\frac{\sum_{i = 1}^{n} |x_i|^{\hat{p}} \ln{|x_i|}}{\sum_{i = 1}^{n} |x_i|^{\hat{p}}} = \frac{\sum_{i = 1}^{n} |x_i|^{2 - \hat{p}} \ln{|x_i|}}{\sum_{i = 1}^{n} |x_i|^{2 - \hat{p}}},
\end{equation*}
i.e., $g(\hat{p}) = g(2 - \hat{p})$, which is a contradiction. Hence, $f(p)$ attains its minimum at $p = 1$.
\end{proof}

Now we prove Observation~\ref{proposition:inner_product_estimation}.

\begin{proof}
Let $\Tilde{\sigma} = ||x||_p^{p / 2} \sqrt{\langle x^{(2 - p)}, y^{(2)} \rangle}$ and $\mathbb{E}_y[y_i^2] = \upsilon$, where $\upsilon$ is a constant. If we fix $\epsilon \Tilde{\sigma} = c_1$ and $1 - \delta = c_2$ in Observation~\ref{prop:general_inner_product}, then the number of queries and samples is $O\left(\frac{\Tilde{\sigma}^2}{c_1^2} \log\frac{1}{1 - c_2}\right) = O\left(\frac{||x||_p^p \langle x^{(2 - p)}, y^{(2)} \rangle}{c_1^2} \log\frac{1}{1 - c_2}\right)$. Then from Observation~\ref{prop:general_inner_product}, we see that reducing $||x||_p^p ||x||_{2 - p}^{2 - p}$ with respect to $p$ allows us to achieve the same precision with smaller number of queries and samples on average. This is because

\begin{align*}
\mathbb{E}_y\big[\text{Var}[||x||_{p}^{p} Z(i)]_{i \sim \mathcal{D}^{(p)}_x}\big] & = ||x||_{p}^{p} \mathbb{E}_y[\langle x^{(2 - p)}, y^{(2)} \rangle] - \mathbb{E}_y[\langle x, y \rangle^2]\\
& = ||x||_{p}^{p} \langle x^{(2 - p)}, \mathbb{E}_y[y^{(2)}] \rangle - \mathbb{E}_y[\langle x, y \rangle^2]\\
& = \upsilon ||x||_p^p ||x||_{2 - p}^{2 - p} - \mathbb{E}_y[\langle x, y \rangle^2]
\end{align*}
and $\mathbb{E}_y[\langle x, y \rangle^2]$ does not depend on $p$. From Lemma~\ref{lemma:attain_minimum}, the optimal value of $p$ is $1$.
\end{proof}

\section{Proof of Observation~\ref{corollary:gaussian_improvement}}
\label{appendix:improvement}

\begin{proof}
Recall that the (average) sample efficiency is proportional to

\begin{equation*}
\mathbb{E}_x\left[\upsilon ||x||_p^p ||x||_{2 - p}^{2 - p} - \mathbb{E}_y[\langle x, y \rangle^2]\right] = \mathbb{E}_x\left[\upsilon \sum_{i = 1}^{n}|x_i|^p \sum_{i = 1}^{n}|x_i|^{2 - p} - \mathbb{E}_y\left[\left(\sum_{i = 1}^{n} x_i y_i\right)^2\right]\right].
\end{equation*}
Furthermore,

\begin{align*}
\mathbb{E}_x\left[\mathbb{E}_y\left[\left(\sum_{i = 1}^{n} x_i y_i\right)^2\right]\right] & = \mathbb{E}_{x, y}\left[\sum_{i = 1}^{n} x_i^2 y_i^2 + 2 \sum_{i < j} x_i x_j y_i y_j \right]\\
& = \sum_{i = 1}^{n}\mathbb{E}_{x}\left[ x_i^2 \right] \mathbb{E}_{y}\left[ y_i^2 \right]\\
& = n \upsilon^2.
\end{align*}
For $p = 2$, we have

\begin{align*}
\mathbb{E}_x\left[\upsilon \sum_{i = 1}^{n}|x_i|^2 \sum_{i = 1}^{n}1 - \mathbb{E}_y[\langle x, y \rangle^2]\right] & = n \upsilon \mathbb{E}_x\left[ \sum_{i = 1}^{n} x_i^2 \right] - n \upsilon^2\\
& = n(n - 1) \upsilon^2.
\end{align*}
For $p = 1$, we have

\begin{align*}
\mathbb{E}_x\left[\upsilon \left(\sum_{i = 1}^{n}|x_i|\right)^2 - \mathbb{E}_y[\langle x, y \rangle^2]\right] & = \upsilon \mathbb{E}_x\left[ \sum_{i = 1}^{n}x_i^2 + \sum_{i \neq j}|x_i| |x_j| \right] - n \upsilon^2\\
& = n \upsilon^2 + n(n - 1)\upsilon \mathbb{E}_x[|x_i|]^2 - n \upsilon^2\\
& = n(n - 1)\upsilon \mathbb{E}_x[|x_i|]^2.
\end{align*}
Therefore, the sample efficiency for $p = 1$ is $\upsilon / \mathbb{E}_x[|x_i|]^2$ times better than that for $p = 2$. For zero mean Gaussian and uniform distributions, one can easily check that $\upsilon / \mathbb{E}_x[|x_i|]^2$ evaluates to $\pi / 2$ and $4 / 3$, respectively.
\end{proof}

\section{Proof of Observation~\ref{prop:general_linear_combination}}
\label{sec:proof_of_proposition_general_linear_combination}

\begin{proof}
$\text{SQ}_{p}(A)$ implies query access to $||A^{(j)}||_p^p$. In the form of rejection sampling, $P$ is the distribution over $i \in [m]$ formed by first sampling $j \in [n]$ with probability proportional to $||x_j A^{(j)}||_p^p$ and then sampling $i \in [m]$ from $\mathcal{D}_{A^{(j)}}^{(p)}$. The target distribution is $Q := \mathcal{D}_{Ax}^{(p)}$. Note that

\begin{align*}
P(i) & = \sum_{j = 1}^{n} \frac{||x_{j} A^{(j)}||_p^p}{\sum_{j' = 1}^{n} ||x_{j'} A^{(j')}||_p^p} \frac{|A_{ij}|^p}{||A^{(j)}||_p^p}\\
& = \frac{\sum_{j = 1}^{n} |x_j A_{ij}|^p}{\sum_{j = 1}^{n} ||x_{j} A^{(j)}||_p^p}.
\end{align*}
The constant $M$ must be chosen such that we can calculate $r_i$ efficiently. We have

\begin{align*}
M & = \frac{1}{P(i)} Q(i) \frac{1}{r_i}\\
& = \frac{\sum_{j = 1}^{n} ||x_{j} A^{(j)}||_p^p}{\sum_{j = 1}^{n} |x_j A_{ij}|^p} \frac{|(Ax)_i|^p}{||Ax||_p^p} \frac{1}{r_i}.
\end{align*}
The terms that do not depend on $i$ are $\sum_{j = 1}^{n} ||x_{j} A^{(j)}||_p^p$ and $||Ax||_p^p$. Therefore, to make $M$ constant,

\begin{equation*}
\frac{|(Ax)_i|^p}{r_i \sum_{j = 1}^{n} |x_j A_{ij}|^p} = \frac{|\sum_{j = 1}^{n} x_j A_{ij}|^p}{r_i \sum_{j = 1}^{n} |x_j A_{ij}|^p}
\end{equation*}
must be constant, i.e.,

\begin{equation*}
r_i = k \frac{|\sum_{j = 1}^{n} x_j A_{ij}|^p}{\sum_{j = 1}^{n} |x_j A_{ij}|^p}
\end{equation*}
where $k$ is a constant. From H\"older's inequality, we can show that

\begin{align}
\label{equation:holder}
n^{p - 1} \sum_{j = 1}^{n} |x_j A_{ij}|^p & = \left\{\left( 
\sum_{j = 1}^n 1^{\frac{p}{p - 1}} \right)^{\frac{p - 1}{p}} \left(\sum_{j = 1}^{n} |x_j A_{ij}|^p\right)^{\frac{1}{p}}\right\}^p \nonumber \\
& \geq \left( \sum_{j = 1}^{n} |x_j A_{ij}| \right)^p \geq \left|\sum_{j = 1}^{n} x_j A_{ij}\right|^p
\end{align}
for $p > 1$ (the inequality trivially holds for $p = 1$; we just can't place $p - 1$ in the denominator). To make $r_i \leq 1$, we choose $k = 1 / n^{p - 1}$ which gives $M = n^{p - 1} \sum_{j = 1}^{n} ||x_{j} A^{(j)}||_p^p / ||Ax||_p^p$. Note that in each iteration, the dominant time complexity is that for sampling $j \in [n]$ and calculating $r_i$, which requires $O(n)$ queries. Since $M$ is the expected number of iterations for a single valid sample, we have our result.
\end{proof}

\section{Proof of Observation~\ref{lemma:positive}}
\label{sec:proof_of_observation_positive}

\begin{proof}
Note that

\begin{equation}
\label{equation:m_p_expression}
M(p) = n^{p - 1} \frac{\sum_{i = 1}^{m} \sum_{j = 1}^{n} |x_{j} A_{ij}|^p}{\sum_{i = 1}^{m} |\sum_{j = 1}^{n} x_{j} A_{ij}|^p}.
\end{equation}
From (\ref{equation:holder}), it immediately follows that $n^{p - 1} \sum_{i = 1}^{m} \sum_{j = 1}^{n} |x_{j} A_{ij}|^p \geq \sum_{i = 1}^{m} |\sum_{j = 1}^{n} x_{j} A_{ij}|^p$ and $M(p) \geq 1$. If $\text{sgn}(A_i)$\hspace{0pt}$= \pm\text{sgn}(x)$, then $|\sum_{j = 1}^{n} x_{j} A_{ij}| = \big|\sum_{j = 1}^{n} \pm|x_{j}||A_{ij}|\big| = \sum_{j = 1}^{n} |x_{j}||A_{ij}|$. Therefore, $M(1) = 1 \leq$\hspace{0pt}$M(p)$. Now consider the extreme case: $A^{(1)}$ is nonzero, $A^{(j)}$ is a zero vector for each $j \in [n] \setminus \{ 1 \}$, and $x_1$ is nonzero. We have $M(p) = n^{p - 1}$ and if $p > 1$, then $M(p) / M(1) = n^{p - 1}$ can get arbitrarily large with increasing $n$.
\end{proof}

\section{Proof of Observation~\ref{lemma:rigorous_proof}}
\label{sec:proof_of_lemma_rigorous_proof}

We make use of the following two facts.

\begin{theorem}[{\citeauthor{goodman1960exact}, \citeyear{goodman1960exact}}]
\label{theorem:product_variance}
Let $X, Y$ be independent random variables with means $\mu_X, \mu_Y$, and variances $\sigma_X^2, \sigma_Y^2$. Then the variance of $XY$ is
\begin{equation*}
\sigma_{XY} = (\sigma_X^2 + \mu_X^2)(\sigma_Y^2 + \mu_Y^2) - \mu_X^2 \mu_Y^2.
\end{equation*}
\end{theorem}

\begin{theorem}[{\citeauthor{winkelbauer2012moments}, \citeyear{winkelbauer2012moments}}]
\label{theorem:absolute_moment}
The central absolute moments of a Gaussian random variable $X \sim \mathcal{N}(\mu, \sigma^2)$ are
\begin{equation*}
\mathbb{E}[|X - \mu|^p] = \sigma^p 2^{p / 2} \frac{\Gamma\left(\frac{p + 1}{2}\right)}{\sqrt{\pi}},
\end{equation*}
where $\Gamma$ is the gamma function.
\end{theorem}

Now we prove Observation~\ref{lemma:rigorous_proof}.

\begin{proof}
Since

\begin{align*}
\frac{M(p)}{n^{p / 2}} & = n^{p / 2 - 1} \frac{\sum_{i = 1}^{m} \sum_{j = 1}^{n} |x_{j} A_{ij}|^p}{\sum_{i = 1}^{m} |\sum_{j = 1}^{n} x_{j} A_{ij}|^p}\\
& = n^{p / 2 - 1} \frac{\sum_{i = 1}^{m} \sum_{j = 1}^{n} |x_{j} A_{ij}|^p}{mn} \frac{mn}{n^{p / 2} \sum_{i = 1}^{m} |\sum_{j = 1}^{n} x_{j} A_{ij} / \sqrt{n}|^p}\\
& = \frac{\sum_{i = 1}^{m} \sum_{j = 1}^{n} |x_{j} A_{ij}|^p}{mn} \frac{m}{\sum_{i = 1}^{m} |\sum_{j = 1}^{n} x_{j} A_{ij} / \sqrt{n}|^p},
\end{align*}
it suffices to show that $\sum_{i = 1}^{m} \sum_{j = 1}^{n} |x_{j} A_{ij}|^p / mn$ and $\sum_{i = 1}^{m} |\sum_{j = 1}^{n} x_{j} A_{ij} / \sqrt{n}|^p / m$ each converge in probability to a constant (which is a function of $p$). For the first term, we have

\begin{align*}
\mathbb{E}\left[\frac{\sum_{i = 1}^{m} \sum_{j = 1}^{n} |x_{j} A_{ij}|^p}{mn}\right] & = \frac{1}{mn} \sum_{j = 1}^{n} \mathbb{E}\left[ |x_{j}|^p \sum_{i = 1}^{m} |A_{ij}|^p \right]\\
& = \frac{1}{mn} \sum_{j = 1}^{n} \mathbb{E}[ |x_{j}|^p ] \mathbb{E}\left[ \sum_{i = 1}^{m} |A_{ij}|^p \right]\\
& = \frac{1}{mn} \sum_{i = 1}^{m} \sum_{j = 1}^{n} \mathbb{E}[ |x_{j}|^p ] \mathbb{E}[ |A_{ij}|^p ]\\
& = \mu_{f, p}^2\\
\end{align*}
and

\begin{align*}
\text{Var}\left[ |x_{j}|^p \sum_{i = 1}^{m} |A_{ij}|^p \right] & = (\sigma_{f, p}^2 + \mu_{f, p}^2) (m \sigma_{f, p}^2 + m^2 \mu_{f, p}^2) - m^2 \mu_{f, p}^4\\
& = m \sigma_{f, p}^2 \{ \sigma_{f, p}^2 + (m + 1)\mu_{f, p}^2 \},
\end{align*}
where the first equality follows from Theorem~\ref{theorem:product_variance} and the independence of the matrix (vector) elements. Summing over $j$ and dividing by $mn$, we have

\begin{align*}
\text{Var}\left[\frac{\sum_{i = 1}^{m} \sum_{j = 1}^{n} |x_{j} A_{ij}|^p}{mn}\right] & = \frac{\sigma_{f, p}^2 \{ \sigma_{f, p}^2 + (m + 1)\mu_{f, p}^2 \}}{mn},
\end{align*}
since $\{ \sum_{j = 1}^{n} |x_{j} A_{ij}|^p \}$ consists of mutually independent random variables. As a result, the term $\sum_{i = 1}^{m} \sum_{j = 1}^{n} |x_{j} A_{ij}|^p / mn$ converges in probability to $\mu_{f, p}^2$ as $n \rightarrow \infty$.

For the second term, note that

\begin{align*}
& \text{Cov}\left( \sum_{j = 1}^{n} x_j A_{ij}, \sum_{j = 1}^{n} x_j A_{i'j} \right)\\
= & \mathbb{E}\left[ \sum_{j, j' = 1}^{n} x_j x_{j'} A_{ij} A_{i'j'} \right] - \mathbb{E}\left[ \sum_{j = 1}^{n} x_j A_{ij} \right] \mathbb{E}\left[ \sum_{j = 1}^{n} x_j A_{i'j} \right]\\
= & 0
\end{align*}
and that any linear combination

\begin{equation*}
\alpha \frac{\sum_{j = 1}^{n} x_j A_{ij}}{\sqrt{n}} + \beta \frac{\sum_{j = 1}^{n} x_j A_{i'j}}{\sqrt{n}} = \frac{\sum_{j = 1}^{n} x_j (\alpha A_{ij} + \beta A_{i'j})}{\sqrt{n}}
\end{equation*}
converges in distribution to $\mathcal{N}(0, (\alpha^2 + \beta^2)\sigma_f^4)$ as $n \rightarrow \infty$. This implies that the joint distribution of $\sum_{j = 1}^{n} x_j A_{ij} / \sqrt{n}$ and $\sum_{j = 1}^{n} x_j A_{i'j} / \sqrt{n}$ converges to the bivariate normal distribution $\mathcal{N}\left( \begin{pmatrix} 0 \\ 0 \\ \end{pmatrix}, \begin{pmatrix} \sigma_f^4 & 0 \\ 0 & \sigma_f^4 \\ \end{pmatrix} \right)$, i.e., the two random variables become independent Gaussians. Consequently, the absolute moments $|\sum_{j = 1}^{n} x_j A_{ij} / \sqrt{n}|^p$ and $|\sum_{j = 1}^{n} x_j A_{i'j} / \sqrt{n}|^p$ also become independent. Averaging over $i$, we conclude that as $m, n \rightarrow \infty$, the variance of $\sum_{i = 1}^{m} |\sum_{j = 1}^{n} x_{j} A_{ij} / \sqrt{n}|^p / m$ converges to $0$ and $\sum_{i = 1}^{m} |\sum_{j = 1}^{n} x_{j} A_{ij} / \sqrt{n}|^p / m$ converges in probability to $\Tilde{\mu}_{f, p}$ due to Theorem~\ref{theorem:absolute_moment}.
\end{proof}

\section{Plots for Observation~\ref{lemma:rigorous_proof}}
\label{section:many_figures_appendix}

\begin{figure}[H]
    \centering
    \begin{minipage}{0.2425\linewidth}
        \includegraphics[width=\linewidth]{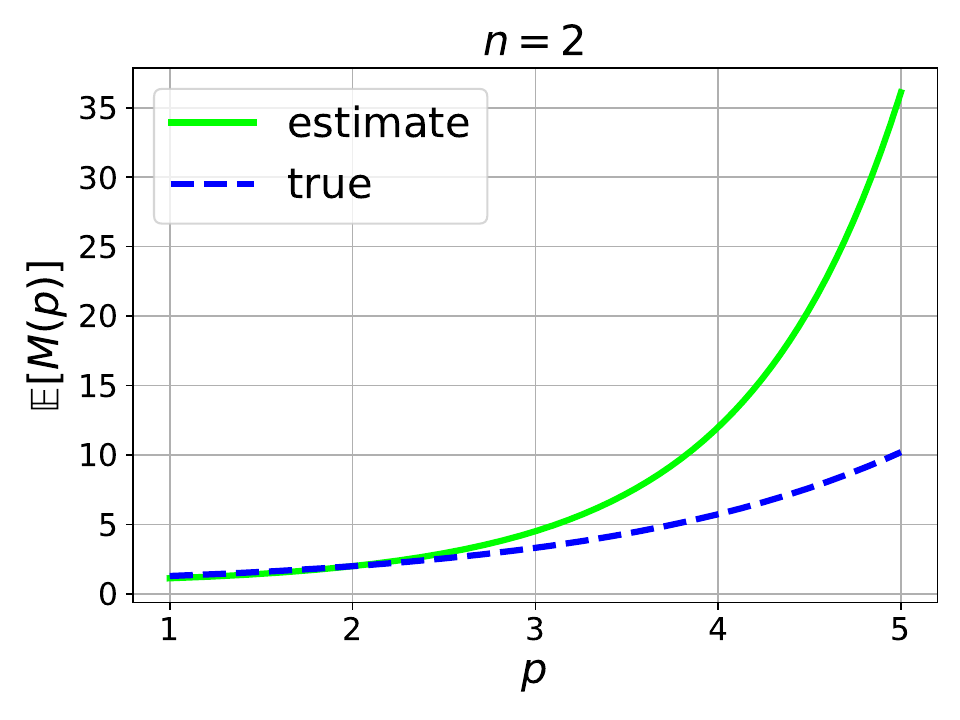}
    \end{minipage}
    \begin{minipage}{0.2425\linewidth}
        \includegraphics[width=\linewidth]{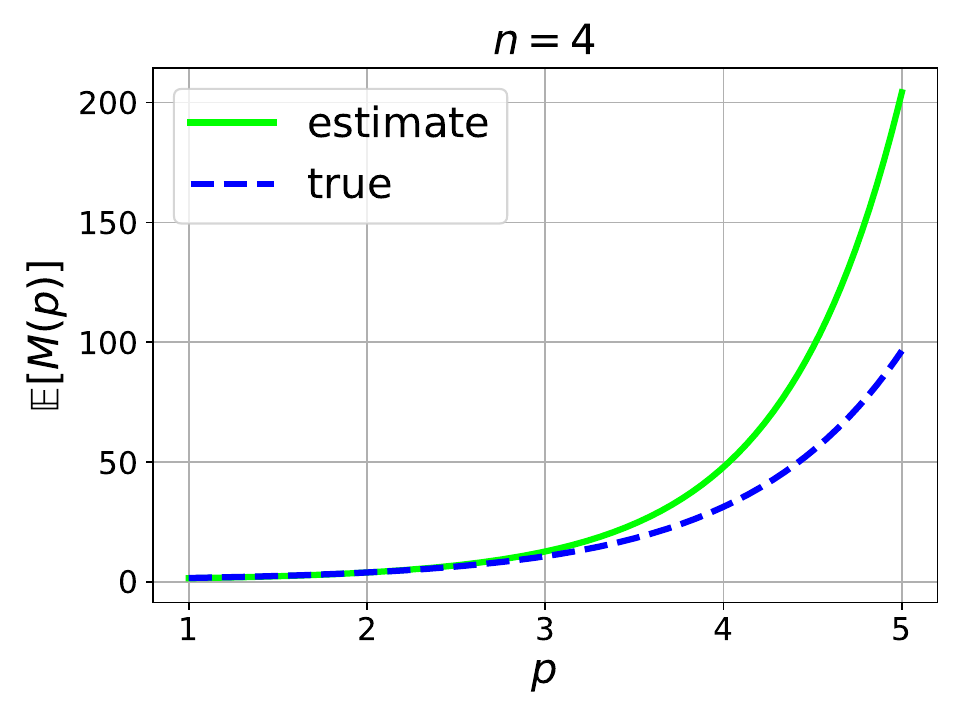}
    \end{minipage}
    \begin{minipage}{0.2425\linewidth}
        \includegraphics[width=\linewidth]{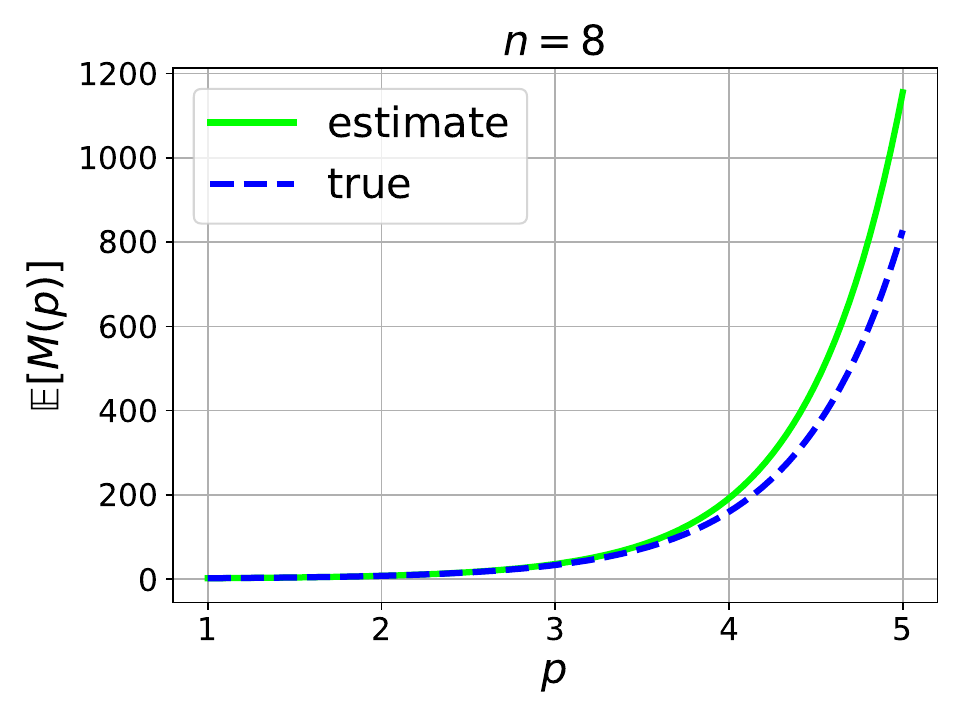}
    \end{minipage}
    \begin{minipage}{0.2425\linewidth}
        \includegraphics[width=\linewidth]{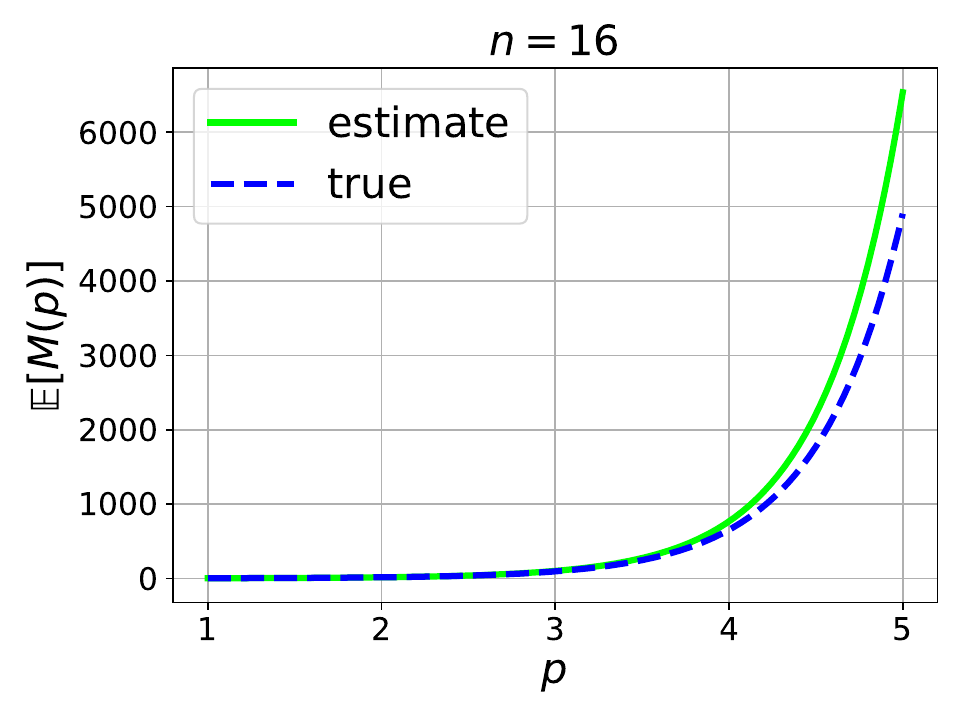}
    \end{minipage}
    
    \begin{minipage}{0.2425\linewidth}
        \includegraphics[width=\linewidth]{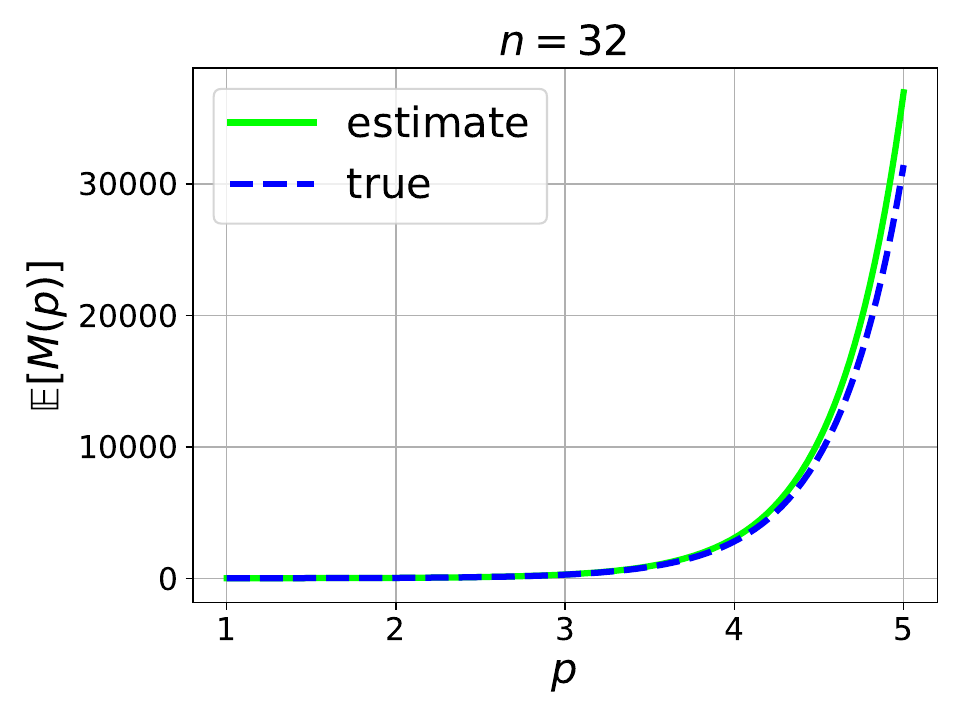}
    \end{minipage}
    \begin{minipage}{0.2425\linewidth}
        \includegraphics[width=\linewidth]{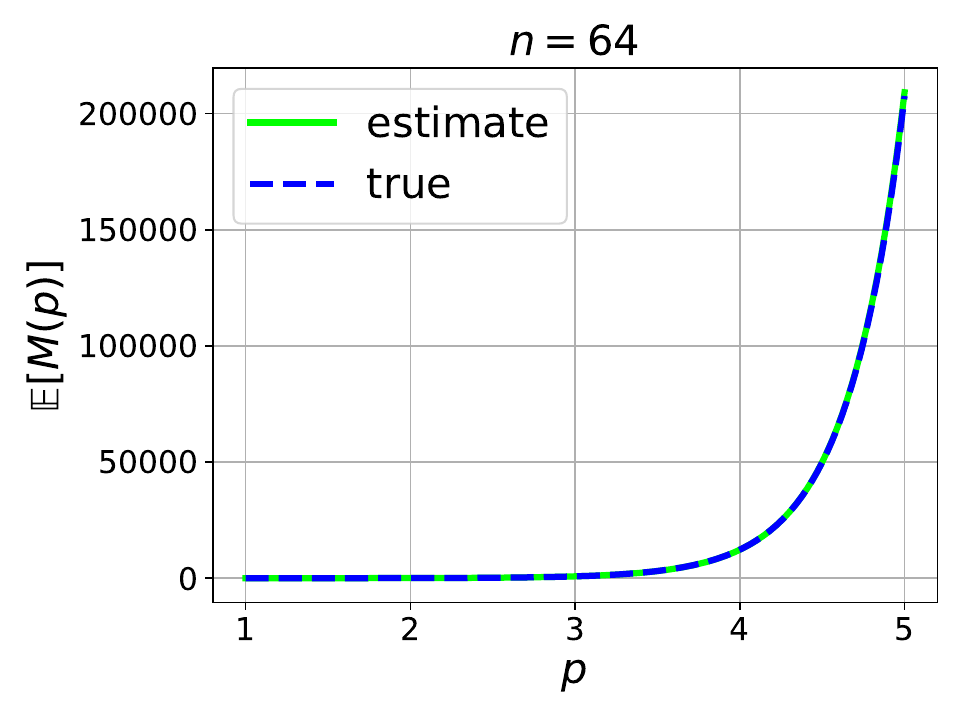}
    \end{minipage}
    \begin{minipage}{0.2425\linewidth}
        \includegraphics[width=\linewidth]{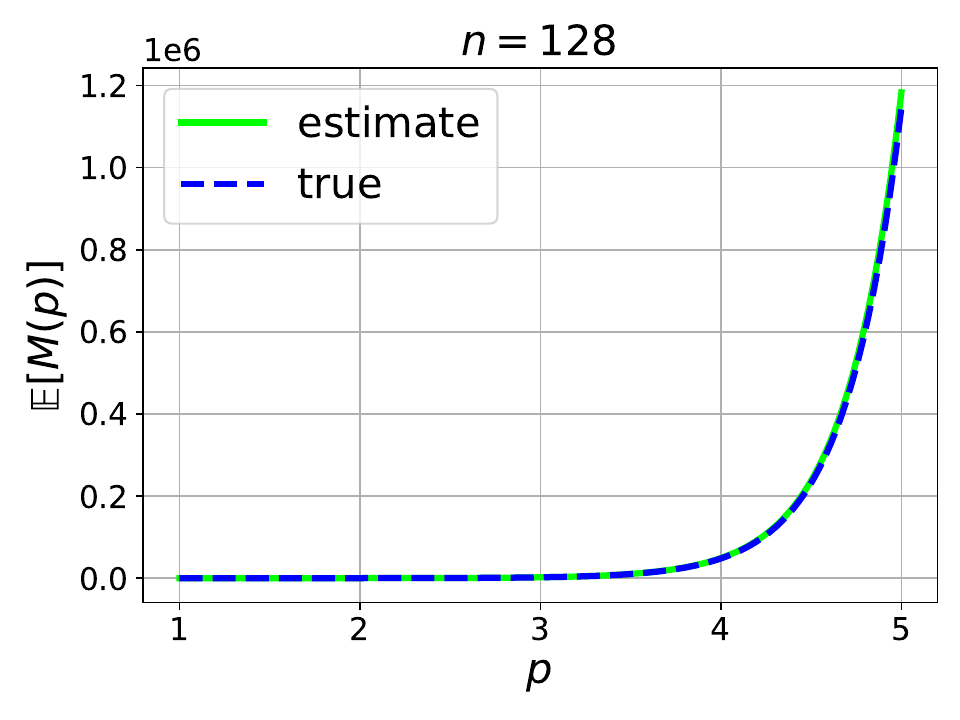}
    \end{minipage}
    \begin{minipage}{0.2425\linewidth}
        \includegraphics[width=\linewidth]{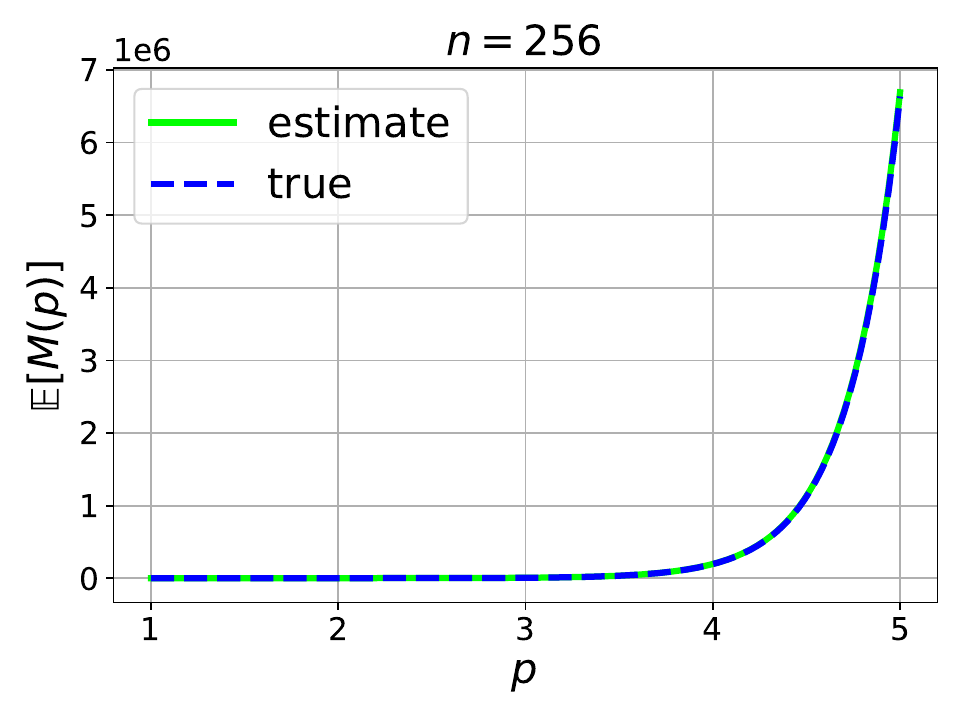}
    \end{minipage}

    \begin{minipage}{0.2425\linewidth}
        \includegraphics[width=\linewidth]{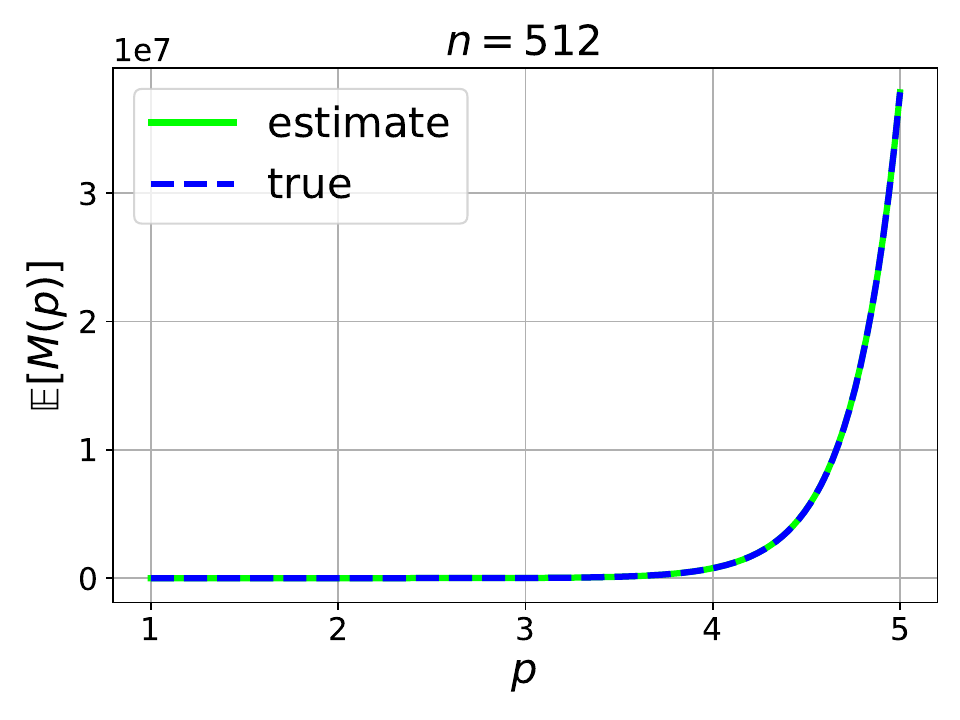}
    \end{minipage}
    \begin{minipage}{0.2425\linewidth}
        \includegraphics[width=\linewidth]{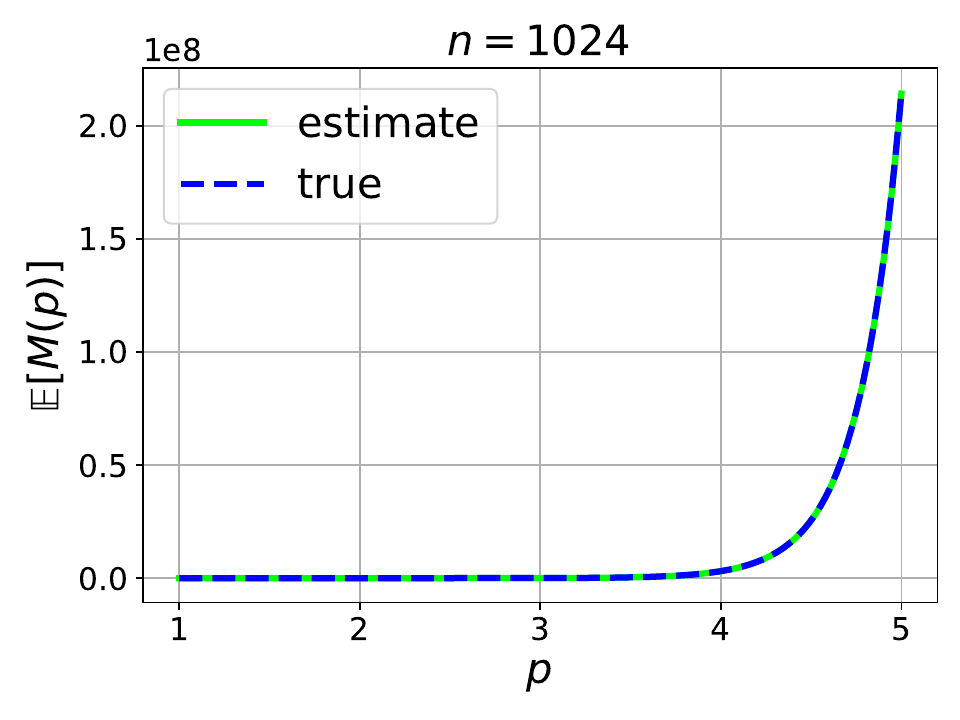}
    \end{minipage}
    \begin{minipage}{0.2425\linewidth}
        \includegraphics[width=\linewidth]{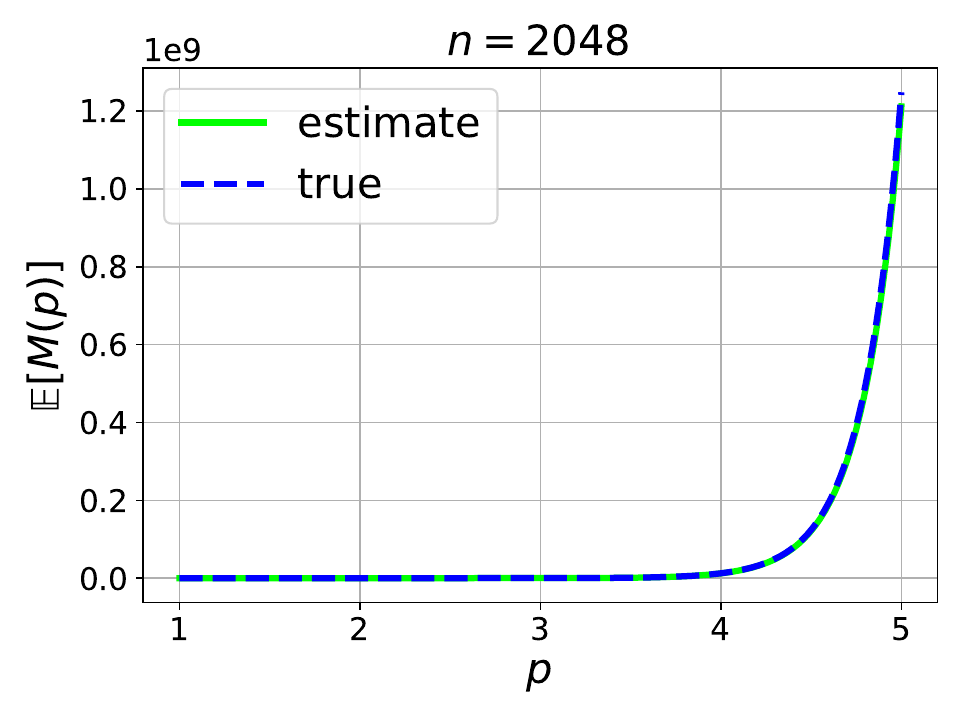}
    \end{minipage}
    \begin{minipage}{0.2425\linewidth}
        \includegraphics[width=\linewidth]{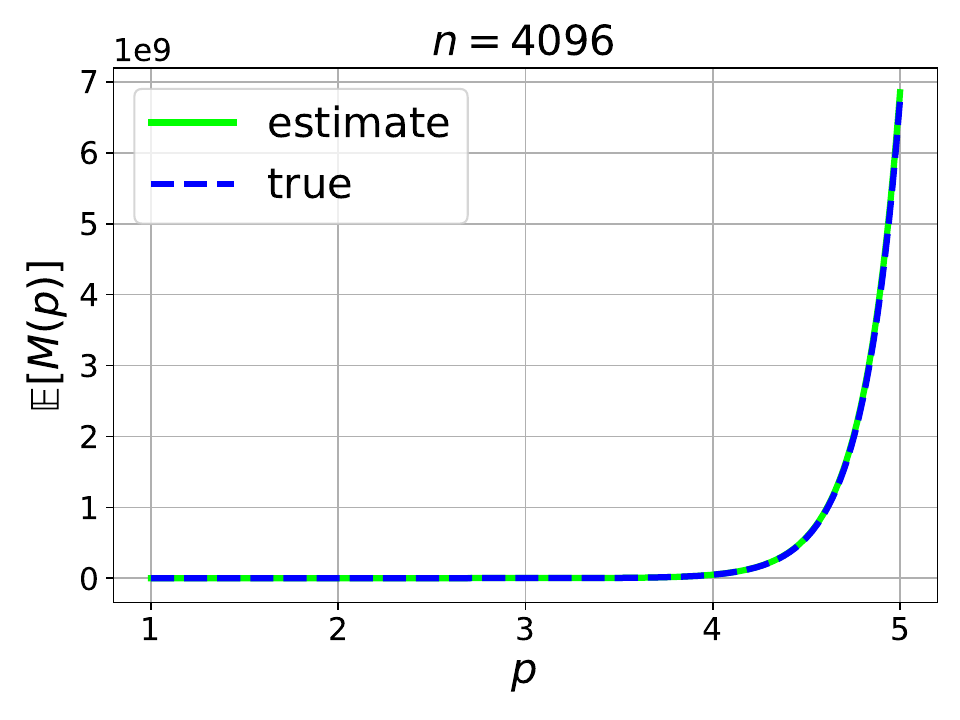}
    \end{minipage}

    \caption{$\mathbb{E}[M(p)]$ vs. $p$ for $f = \mathcal{N}(0, 1)$, $m = 1024$, and $n \in \{ 
2, 4, \cdots , 4096 \}$. Assuming $n \rightarrow \infty$ gives a positively biased estimate for small $n$.}
    \label{fig:many_figures1}
\end{figure}

\begin{figure}[H]
    \centering
    \begin{minipage}{0.2425\linewidth}
        \includegraphics[width=\linewidth]{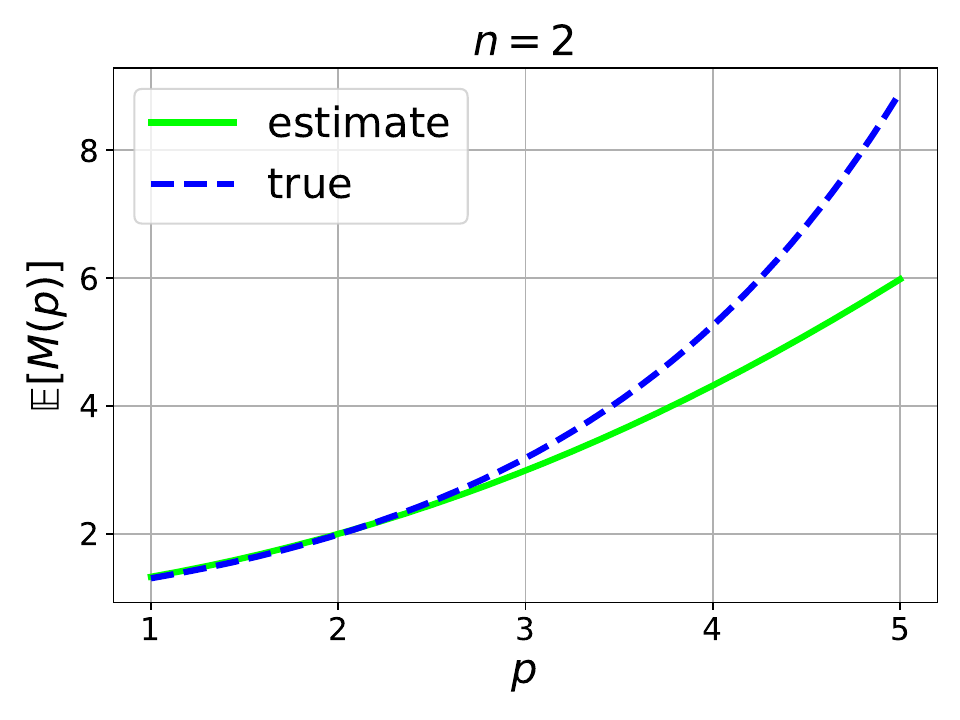}
    \end{minipage}
    \begin{minipage}{0.2425\linewidth}
        \includegraphics[width=\linewidth]{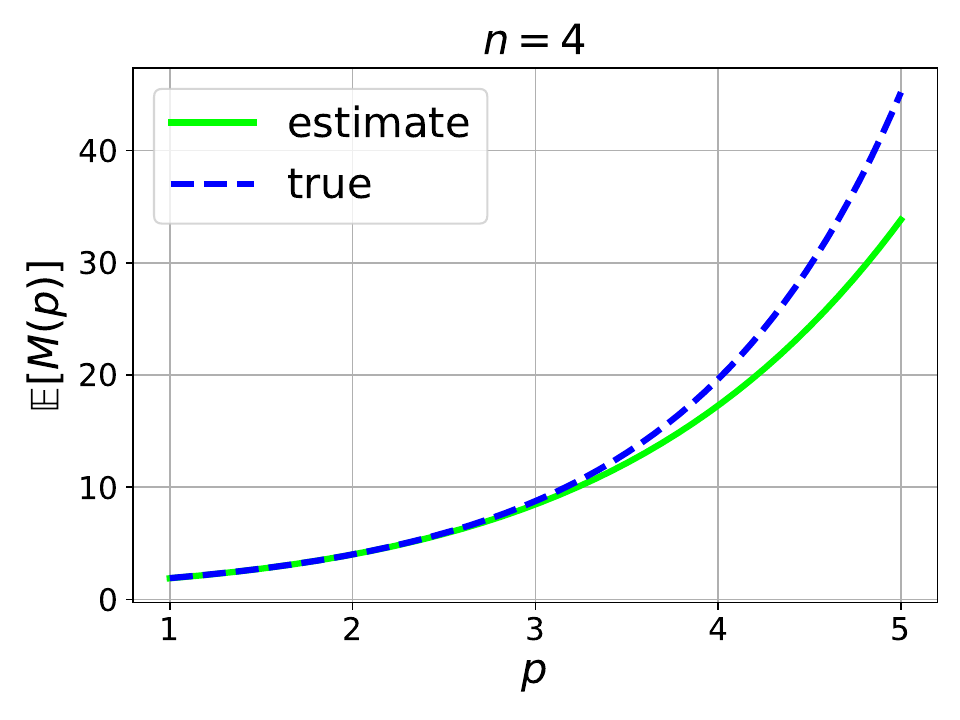}
    \end{minipage}
    \begin{minipage}{0.2425\linewidth}
        \includegraphics[width=\linewidth]{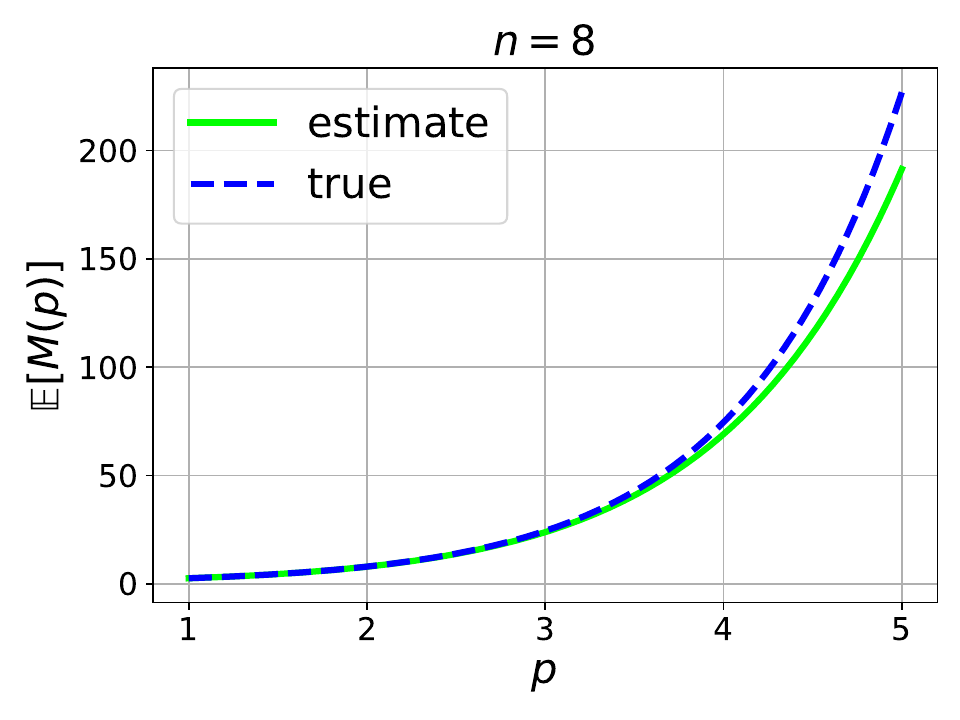}
    \end{minipage}
    \begin{minipage}{0.2425\linewidth}
        \includegraphics[width=\linewidth]{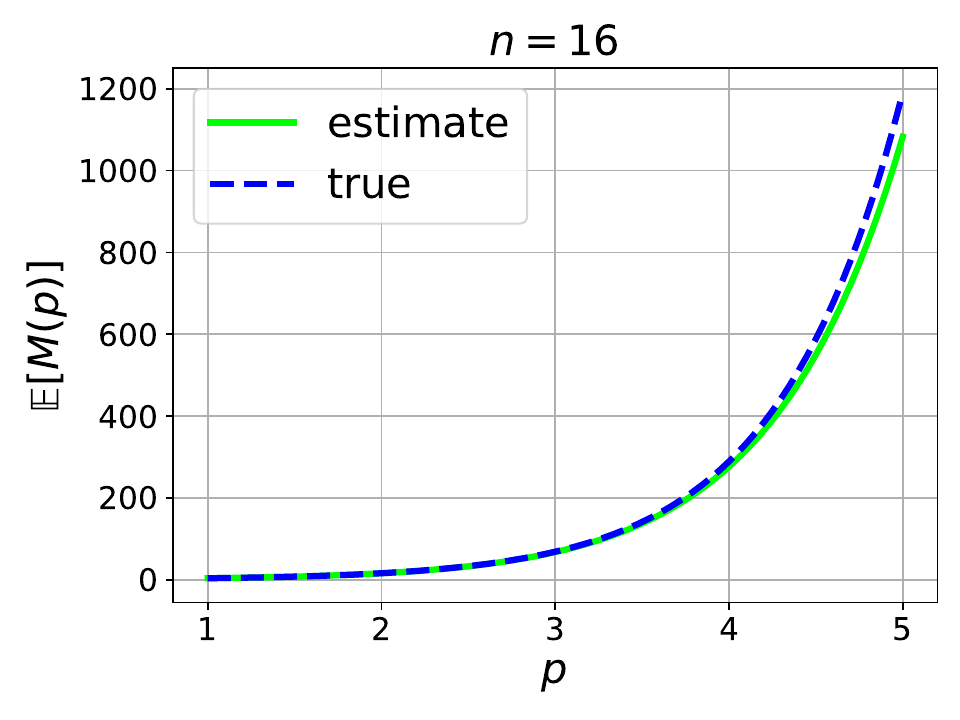}
    \end{minipage}
    
    \begin{minipage}{0.2425\linewidth}
        \includegraphics[width=\linewidth]{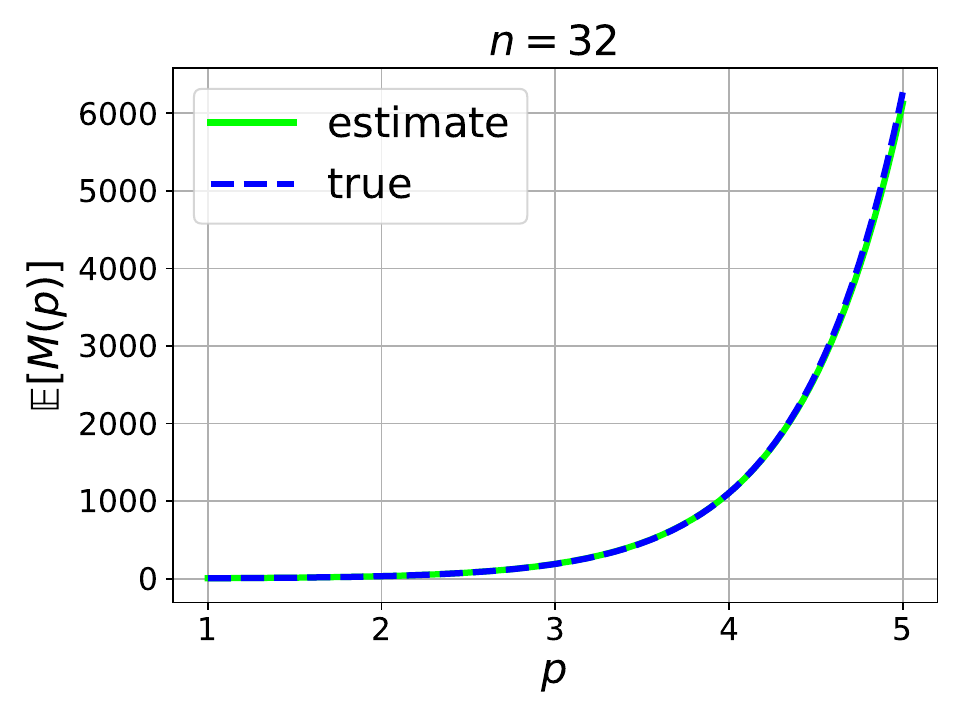}
    \end{minipage}
    \begin{minipage}{0.2425\linewidth}
        \includegraphics[width=\linewidth]{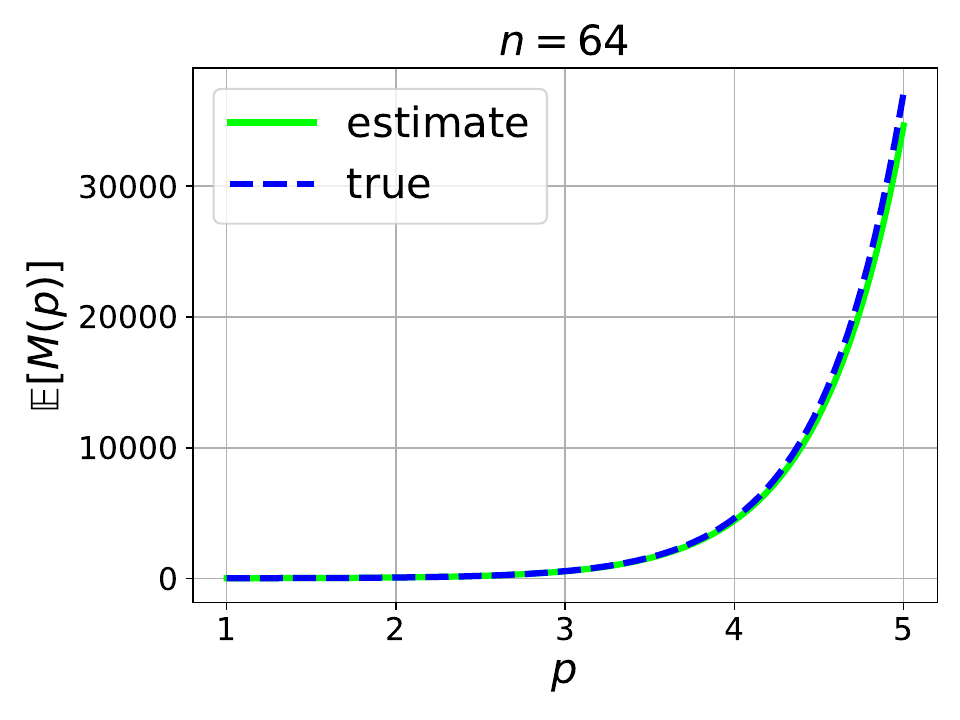}
    \end{minipage}
    \begin{minipage}{0.2425\linewidth}
        \includegraphics[width=\linewidth]{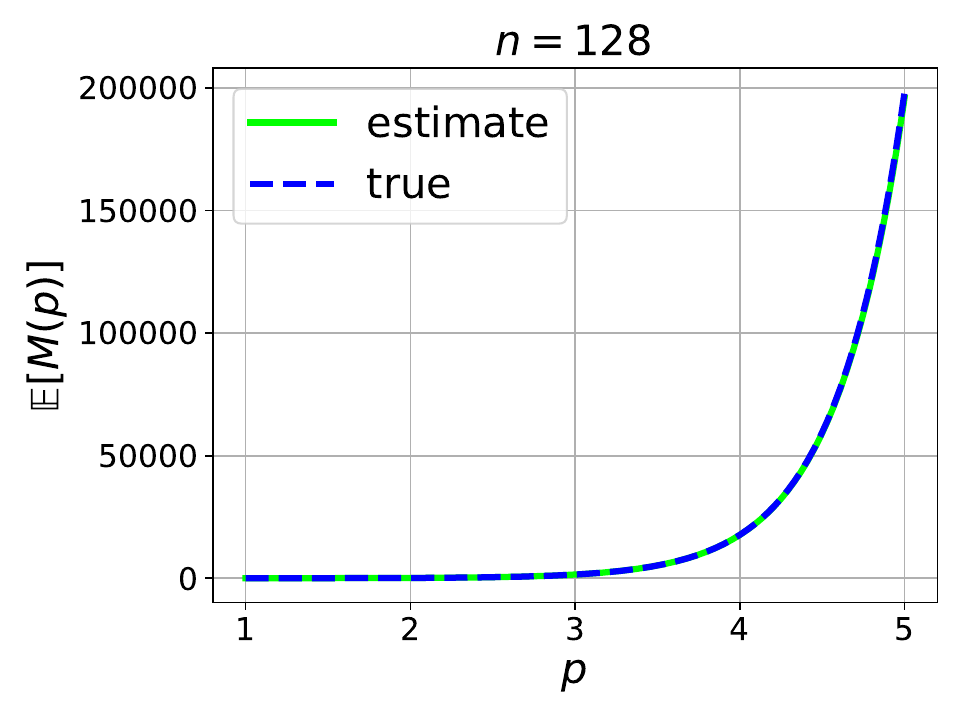}
    \end{minipage}
    \begin{minipage}{0.2425\linewidth}
        \includegraphics[width=\linewidth]{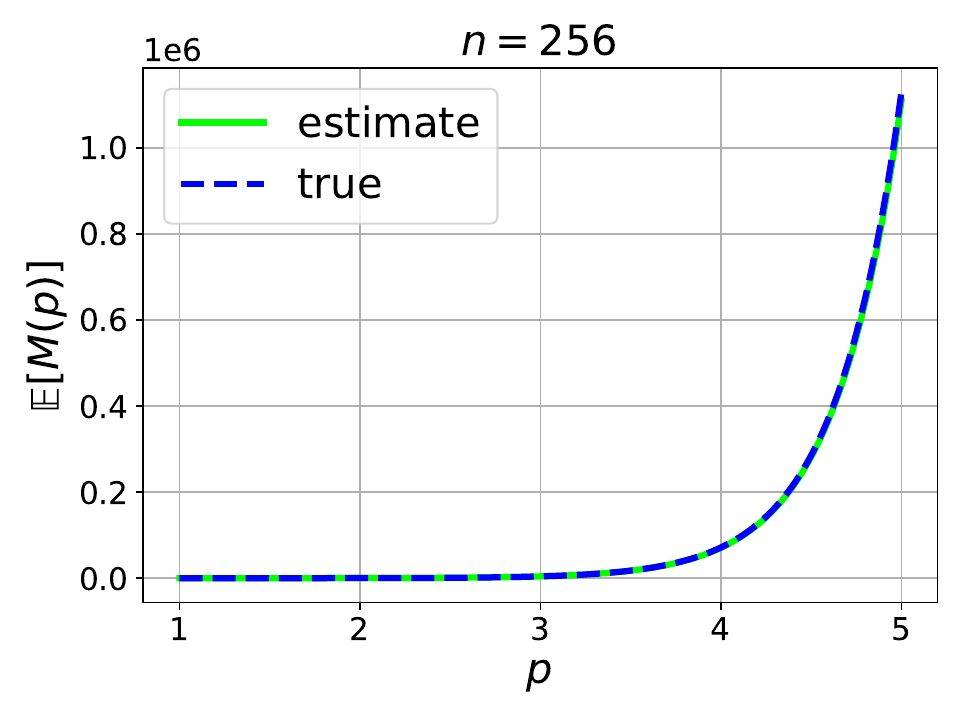}
    \end{minipage}
    
    \begin{minipage}{0.2425\linewidth}
        \includegraphics[width=\linewidth]{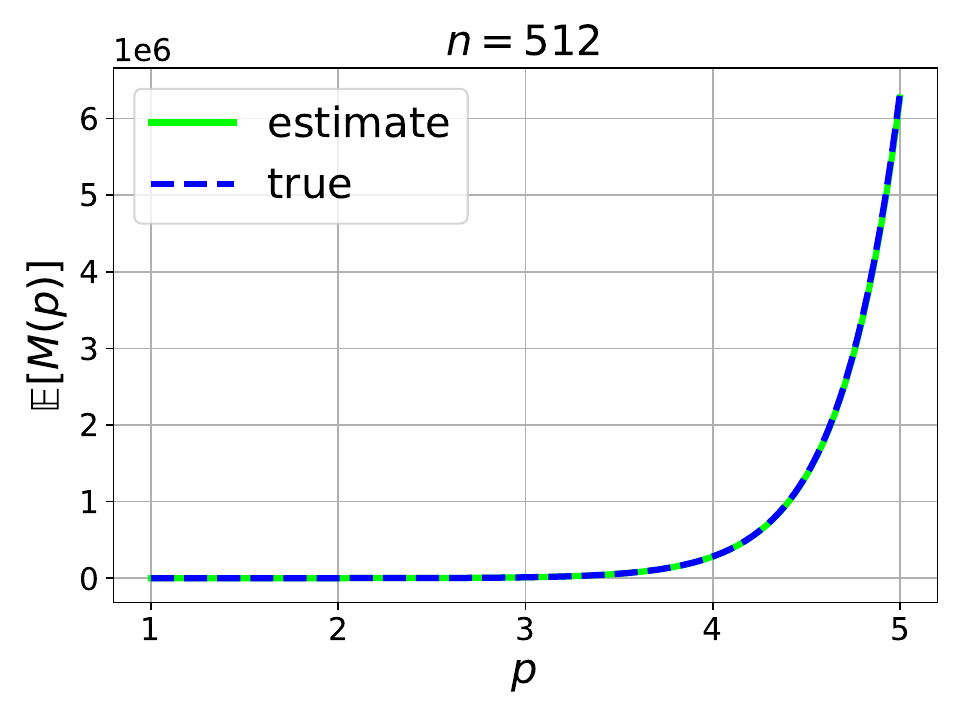}
    \end{minipage}
    \begin{minipage}{0.2425\linewidth}
        \includegraphics[width=\linewidth]{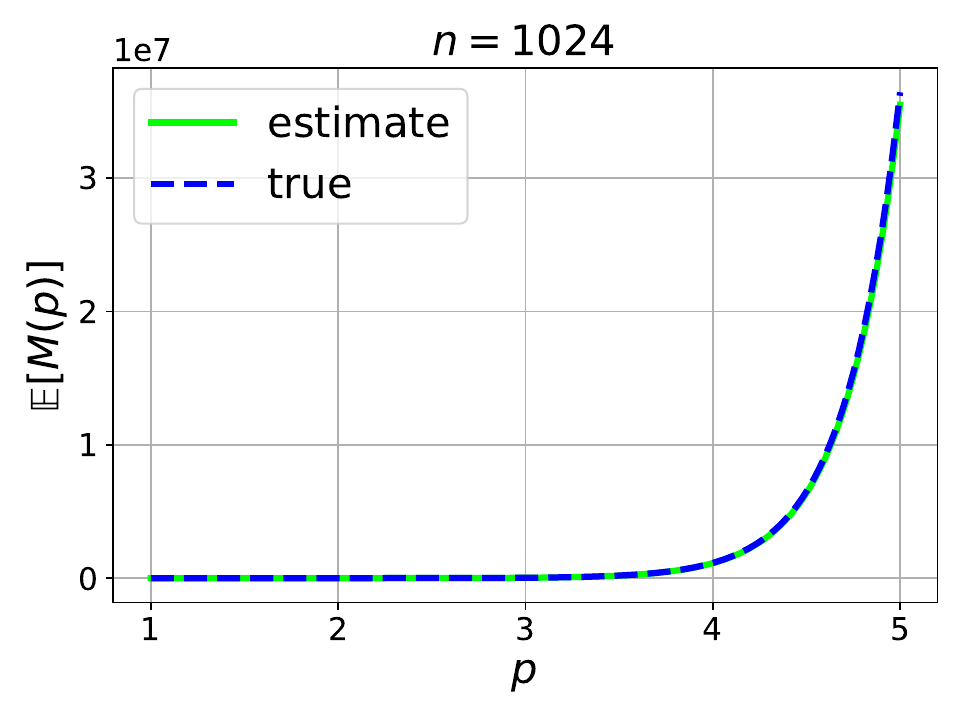}
    \end{minipage}
    \begin{minipage}{0.2425\linewidth}
        \includegraphics[width=\linewidth]{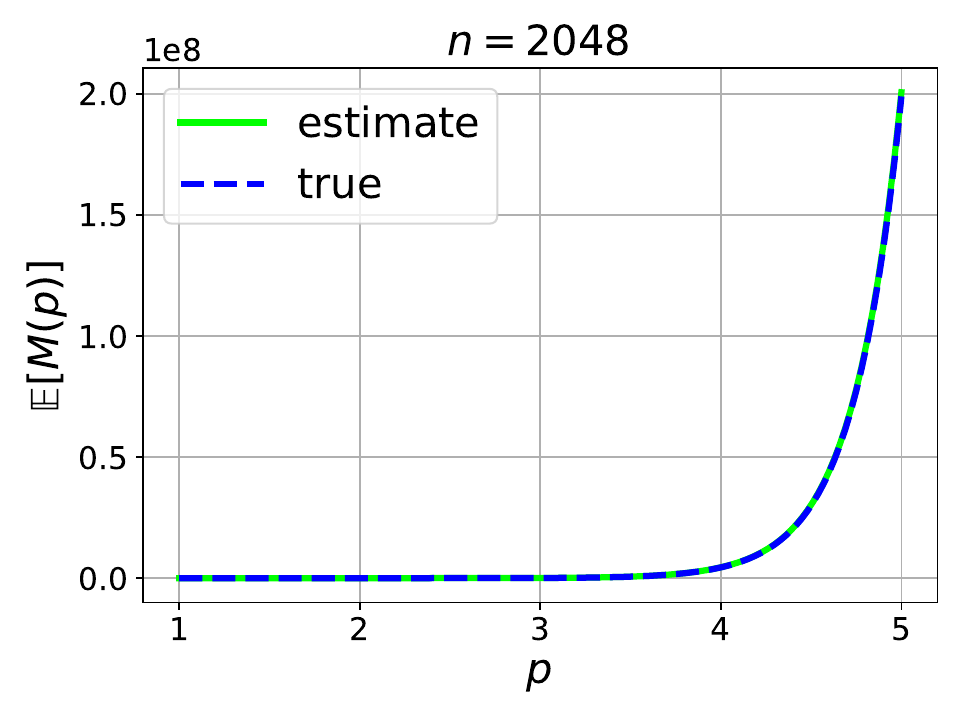}
    \end{minipage}
    \begin{minipage}{0.2425\linewidth}
        \includegraphics[width=\linewidth]{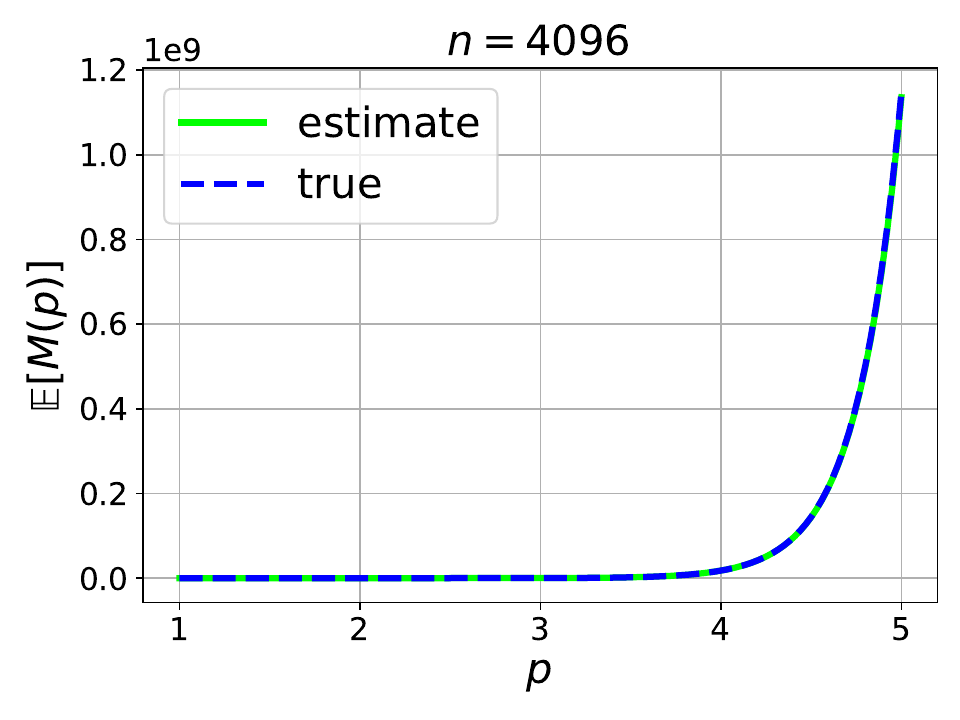}
    \end{minipage}

    \caption{$\mathbb{E}[M(p)]$ vs. $p$ for $f = \mathcal{U}_{(-1, 1)}$, $m = 1024$, and $n \in \{ 
2, 4, \cdots , 4096 \}$. Assuming $n \rightarrow \infty$ gives a negatively biased estimate for small $n$.}
    \label{fig:many_figures2}
\end{figure}

\section{Derivation of (\ref{equation:Z_upper_bound})}
\label{appendix:derive_Z_upper_bound}

\begin{align*}
Z & = \sum_{\textbf{k} \in \{0, 1\}^n} \frac{1}{n \sqrt{d}} |n - 2|\textbf{k}|| + \binom{n}{2}\frac{2}{n\sqrt{d}} 2^{n - 1}\\
& \leq \sqrt{\left(\sum_{\textbf{k} \in \{0, 1\}^n} \frac{(n - 2|\textbf{k}|)^2}{n^2 d}\right) \left(\sum_{\textbf{k} \in \{0, 1\}^n} 1\right)}  + \frac{n(n - 1)}{2}\frac{d}{n \sqrt{d}}\\
& = \sqrt{\frac{d}{n}} + \frac{(n - 1) \sqrt{d}}{2}\\
& = \left(\frac{n}{2} + \frac{1}{\sqrt{n}} - \frac{1}{2}\right) \sqrt{d}
\end{align*}

\section{Derivation of (\ref{equation:binomial_identity})}
\label{appendix:binomial}

\begin{align*}
\sum_{w = 0}^{n} \binom{n}{w} |n - 2w| & = \sum_{w = 0}^{\lceil n / 2 \rceil - 1} \binom{n}{w} |n - 2w| + \sum_{w = \lfloor n / 2 \rfloor + 1}^{n} \binom{n}{w} |n - 2w|\\
& = n \left( \sum_{w = 0}^{\lceil n / 2 \rceil - 1} \binom{n}{w} - \sum_{w = \lfloor n / 2 \rfloor + 1}^{n} \binom{n}{w} \right)\\
& + 2 \left( \sum_{w = \lfloor n / 2 \rfloor + 1}^{n} w\binom{n}{w} - \sum_{w = 0}^{\lceil n / 2 \rceil - 1} w\binom{n}{w} \right)\\
& = 2 \left( \sum_{w = \lfloor n / 2 \rfloor + 1}^{n} w\binom{n}{w} - \sum_{w = 0}^{\lceil n / 2 \rceil - 1} w\binom{n}{w} \right)\\
& = 2 \left( \sum_{w = \lfloor n / 2 \rfloor + 1}^{n} n\binom{n - 1}{w - 1} - \sum_{w = 1}^{\lceil n / 2 \rceil - 1} n\binom{n - 1}{w - 1} \right)\\
& = 2n \left( \sum_{w = \lfloor n / 2 \rfloor}^{n - 1} \binom{n - 1}{w} - \sum_{w = 0}^{\lceil n / 2 \rceil - 2} \binom{n - 1}{w} \right)\\
& = 2n \binom{n - 1}{\lfloor n / 2 \rfloor}
\end{align*}

\section{Proof of Proposition~\ref{proposition:class_of_states}}
\label{sec:proof_of_proposition_class_of_states}

\begin{proof}
Suppose we use $N_i$ measurement outcomes for each $1 \leq i \leq l$. For $\text{SQ}_2$ sampling, $\Tilde{Y}$ is defined as
\begin{equation*}
\Tilde{Y} = \sum_{i = 1}^l \sum_{j = 1}^{N_i} \frac{1}{l N_i \sqrt{d} \chi_{\rho (n)}(k_i)}M_{ij}.
\end{equation*}
From Hoeffding's inequality, it can be shown that $F(\rho, \sigma)$ lies in the range $[\Tilde{Y} - 2\epsilon, \Tilde{Y} + 2\epsilon]$ with probability $\geq 1 - 2\delta$, where the upper bound on the total number of measurements is
\begin{equation*}
\approx \frac{1}{\alpha(n)^2} \frac{2 \log (2 / \delta)}{\epsilon^2}.
\end{equation*}
For $\text{SQ}_1$ sampling, $\Tilde{Y}$ is defined as
\begin{equation*}
\Tilde{Y} = \sum_{i = 1}^l \sum_{j = 1}^{N_i} \frac{Z \text{sgn} (\chi_{\rho (n)}(k_i))}{l N_i \sqrt{d}}M_{ij}.
\end{equation*}
From Hoeffding's inequality, it can be shown that $F(\rho, \sigma)$ lies in the range $[\Tilde{Y} - 2\epsilon, \Tilde{Y} + 2\epsilon]$ with probability $\geq 1 - 2\delta$, where the upper bound on the total number of measurements is
\begin{equation*}
\approx \frac{Z^2}{d} \frac{2 \log (2 / \delta)}{\epsilon^2}.
\end{equation*}
It suffices to show that using $\text{SQ}_1$ sampling improves this bound:
\begin{equation*}
\frac{Z}{\sqrt{d}} \leq \frac{1}{\alpha (n)} \iff Z \leq \frac{\sqrt{d}}{\alpha (n)},
\end{equation*}
where $Z \equiv \sum_k |\chi_{\rho (n)}(k)|$. But by definition,
\begin{equation}
\label{equation:general_proof_num_nonzero}
|\chi_{\rho (n)} (k)| \geq \frac{\alpha (n)}{\sqrt{d}} \quad \text{and} \quad \chi_{\rho (n)} (k)^2 \geq \frac{\alpha (n)^2}{d}.
\end{equation}
Let $K_{+} \equiv \{ k \mid \chi_{\rho (n)} (k)^2 > 0 \}$ denote the set of indices for nonzero probabilities. From (\ref{equation:general_proof_num_nonzero}) and $\sum_k \chi_{\rho (n)} (k)^2 = 1$, it is clear that $|K_{+}| \leq d / \alpha (n)^2$. Therefore,

\begin{equation*}
Z = \sum_{k \in K_{+}} |\chi_{\rho (n)}(k)| \leq \sqrt{\left( \sum_{k \in K_{+}} \chi_{\rho (n)}(k)^2 \right) \left( \sum_{k \in K_{+}} 1 \right)} = \sqrt{|K_{+}|} \leq \frac{\sqrt{d}}{\alpha (n)}.
\end{equation*}
\end{proof}

\subsection*{Author Contributions}
Hyunho Cha: conceptualization and writing. Sunbeom Jeong: experimental assistance. Jungwoo Lee: proofreading.

\subsection*{Data Availability}
No datasets were generated or analyzed during the current study.

\section*{Declarations}
The work is original and unpublished and has not been submitted for publication elsewhere.

\subsection*{Conflict of Interest}
The authors declare no competing interests.





\end{appendices}

\bibliography{sn-article}

@inproceedings{tang2019quantum,
  title={A quantum-inspired classical algorithm for recommendation systems},
  author={Tang, Ewin},
  booktitle={Proceedings of the 51st annual ACM SIGACT symposium on theory of computing},
  pages={217--228},
  year={2019},
howpublished = {\url{https://doi.org/10.1145/3313276.3316310}}
}

@article{winkelbauer2012moments,
  title={Moments and absolute moments of the normal distribution},
  author={Winkelbauer, Andreas},
  journal={arXiv preprint arXiv:1209.4340},
  year={2012}
}

@article{goodman1960exact,
  title={On the exact variance of products},
  author={Goodman, Leo A},
  journal={Journal of the American statistical association},
  volume={55},
  number={292},
  pages={708--713},
  year={1960},
  publisher={Taylor \& Francis}
}

@article{flammia2011direct,
  title={Direct fidelity estimation from few Pauli measurements},
  author={Flammia, Steven T and Liu, Yi-Kai},
  journal={Physical review letters},
  volume={106},
  number={23},
  pages={230501},
  year={2011},
  publisher={APS},
doi = {10.1103/PhysRevLett.106.230501}
}

@article{ronnow2014defining,
  title={Defining and detecting quantum speedup},
  author={R{\o}nnow, Troels F and Wang, Zhihui and Job, Joshua and Boixo, Sergio and Isakov, Sergei V and Wecker, David and Martinis, John M and Lidar, Daniel A and Troyer, Matthias},
  journal={science},
  volume={345},
  number={6195},
  pages={420--424},
  year={2014},
  publisher={American Association for the Advancement of Science},
doi={10.1126/science.1252319}
}

@article{childs2018toward,
  title={Toward the first quantum simulation with quantum speedup},
  author={Childs, Andrew M and Maslov, Dmitri and Nam, Yunseong and Ross, Neil J and Su, Yuan},
  journal={Proceedings of the National Academy of Sciences},
  volume={115},
  number={38},
  pages={9456--9461},
  year={2018},
  publisher={National Acad Sciences},
doi={10.1073/pnas.1801723115}
}

@article{montanaro2015quantum,
  title={Quantum speedup of Monte Carlo methods},
  author={Montanaro, Ashley},
  journal={Proceedings of the Royal Society A: Mathematical, Physical and Engineering Sciences},
  volume={471},
  number={2181},
  pages={20150301},
  year={2015},
  publisher={The Royal Society},
doi={10.1098/rspa.2015.0301}
}

@article{kerenidis2016quantum,
  title={Quantum recommendation systems},
  author={Kerenidis, Iordanis and Prakash, Anupam},
  journal={arXiv preprint arXiv:1603.08675},
  year={2016}
}

@article{shor1999polynomial,
  title={Polynomial-time algorithms for prime factorization and discrete logarithms on a quantum computer},
  author={Shor, Peter W},
  journal={SIAM review},
  volume={41},
  number={2},
  pages={303--332},
  year={1999},
  publisher={SIAM}
}

@article{deutsch1992rapid,
  title={Rapid solution of problems by quantum computation},
  author={Deutsch, David and Jozsa, Richard},
  journal={Proceedings of the Royal Society of London. Series A: Mathematical and Physical Sciences},
  volume={439},
  number={1907},
  pages={553--558},
  year={1992},
  publisher={The Royal Society London}
}

@article{tang2021quantum,
  title={Quantum principal component analysis only achieves an exponential speedup because of its state preparation assumptions},
  author={Tang, Ewin},
  journal={Physical Review Letters},
  volume={127},
  number={6},
  pages={060503},
  year={2021},
  publisher={APS},
doi={10.1103/PhysRevLett.127.060503}
}

@article{chia2022sampling,
  title={Sampling-based sublinear low-rank matrix arithmetic framework for dequantizing quantum machine learning},
  author={Chia, Nai-Hui and Gily{\'e}n, Andr{\'a}s Pal and Li, Tongyang and Lin, Han-Hsuan and Tang, Ewin and Wang, Chunhao},
  journal={Journal of the ACM},
  volume={69},
  number={5},
  pages={1--72},
  year={2022},
  publisher={ACM New York, NY},
doi={10.1145/3549524}
}

@article{gilyen2018quantum,
  title={Quantum-inspired low-rank stochastic regression with logarithmic dependence on the dimension},
  author={Gily{\'e}n, Andr{\'a}s and Lloyd, Seth and Tang, Ewin},
  journal={arXiv preprint arXiv:1811.04909},
  year={2018}
}

@article{leone2023nonstabilizerness,
  title={Nonstabilizerness determining the hardness of direct fidelity estimation},
  author={Leone, Lorenzo and Oliviero, Salvatore FE and Hamma, Alioscia},
  journal={Physical Review A},
  volume={107},
  number={2},
  pages={022429},
  year={2023},
  publisher={APS},
doi={10.1103/PhysRevA.107.022429}
}

@article{zhang2021direct,
  title={Direct fidelity estimation of quantum states using machine learning},
  author={Zhang, Xiaoqian and Luo, Maolin and Wen, Zhaodi and Feng, Qin and Pang, Shengshi and Luo, Weiqi and Zhou, Xiaoqi},
  journal={Physical Review Letters},
  volume={127},
  number={13},
  pages={130503},
  year={2021},
  publisher={APS},
doi={10.1103/PhysRevLett.127.130503}
}

@article{kloek1978bayesian,
  title={Bayesian estimates of equation system parameters: an application of integration by Monte Carlo},
  author={Kloek, Teun and Van Dijk, Herman K},
  journal={Econometrica: Journal of the Econometric Society},
  pages={1--19},
  year={1978},
  publisher={JSTOR}
}

@techreport{goertzel1949quota,
  title={Quota sampling and importance functions in stochastic solution of particle problems},
  author={Goertzel, Gerald},
  year={1949}
}

@article{kahn1951estimation,
  title={Estimation of particle transmission by random sampling},
  author={Kahn, Herman and Harris, Theodore E},
  journal={National Bureau of Standards applied mathematics series},
  volume={12},
  pages={27--30},
  year={1951},
  publisher={BibSonomy}
}

@article{huang2020predicting,
  title={Predicting many properties of a quantum system from very few measurements},
  author={Huang, Hsin-Yuan and Kueng, Richard and Preskill, John},
  journal={Nature Physics},
  volume={16},
  number={10},
  pages={1050--1057},
  year={2020},
  publisher={Nature Publishing Group UK London},
doi={10.1038/s41567-020-0932-7}
}

@article{jozsa1994fidelity,
  title={Fidelity for mixed quantum states},
  author={Jozsa, Richard},
  journal={Journal of modern optics},
  volume={41},
  number={12},
  pages={2315--2323},
  year={1994},
  publisher={Taylor \& Francis},
doi={10.1080/09500349414552171}
}

@book{nielsen2010quantum,
  title={Quantum computation and quantum information},
  author={Nielsen, Michael A and Chuang, Isaac L},
  year={2010},
  publisher={Cambridge university press}
}

@book{wilde2013quantum,
  title={Quantum information theory},
  author={Wilde, Mark M},
  year={2013},
  publisher={Cambridge university press},
howpublished = {\url{https://doi.org/10.1017/CBO9781139525343}}
}

@article{baldwin2023efficiently,
  title={Efficiently computing the Uhlmann fidelity for density matrices},
  author={Baldwin, Andrew J and Jones, Jonathan A},
  journal={Physical Review A},
  volume={107},
  number={1},
  pages={012427},
  year={2023},
  publisher={APS},
doi={10.1103/PhysRevA.107.012427}
}

@article{liang2019quantum,
  title={Quantum fidelity measures for mixed states},
  author={Liang, Yeong-Cherng and Yeh, Yu-Hao and Mendon{\c{c}}a, Paulo EMF and Teh, Run Yan and Reid, Margaret D and Drummond, Peter D},
  journal={Reports on Progress in Physics},
  volume={82},
  number={7},
  pages={076001},
  year={2019},
  publisher={IOP Publishing},
doi={10.1088/1361-6633/ab1ca4}
}

@article{arrazola2020quantum,
  title={Quantum-inspired algorithms in practice},
  author={Arrazola, Juan Miguel and Delgado, Alain and Bardhan, Bhaskar Roy and Lloyd, Seth},
  journal={Quantum},
  volume={4},
  pages={307},
  year={2020},
  publisher={Verein zur F{\"o}rderung des Open Access Publizierens in den Quantenwissenschaften},
doi={10.22331/q-2020-08-13-307}
}

@article{casella2004generalized,
  title={Generalized accept-reject sampling schemes},
  author={Casella, George and Robert, Christian P and Wells, Martin T},
  journal={Lecture notes-monograph series},
  pages={342--347},
  year={2004},
  publisher={JSTOR}
}

@article{neal2003slice,
  title={Slice sampling},
  author={Neal, Radford M},
  journal={The annals of statistics},
  volume={31},
  number={3},
  pages={705--767},
  year={2003},
  publisher={Institute of Mathematical Statistics}
}

@article{forsythe1972neumann,
  title={Von Neumann’s comparison method for random sampling from the normal and other distributions},
  author={Forsythe, George E},
  journal={Mathematics of Computation},
  volume={26},
  number={120},
  pages={817--826},
  year={1972}
}

@article{legault2019accounting,
  title={Accounting for environmental change in continuous-time stochastic population models},
  author={Legault, Geoffrey and Melbourne, Brett A},
  journal={Theoretical Ecology},
  volume={12},
  number={1},
  pages={31--48},
  year={2019},
  publisher={Springer},
doi={10.1007/s12080-018-0386-z}
}

@book{thomopoulos2012essentials,
  title={Essentials of Monte Carlo simulation: Statistical methods for building simulation models},
  author={Thomopoulos, Nick T},
  year={2012},
  publisher={Springer Science \& Business Media}
}

@incollection{greenberger1989going,
  title={Going beyond Bell’s theorem},
  author={Greenberger, Daniel M and Horne, Michael A and Zeilinger, Anton},
  booktitle={Bell’s theorem, quantum theory and conceptions of the universe},
  pages={69--72},
  year={1989},
  publisher={Springer}
}

@article{mermin1990quantum,
  title={Quantum mysteries revisited},
  author={Mermin, N David},
  journal={Am. J. Phys},
  volume={58},
  number={8},
  pages={731--734},
  year={1990}
}

@article{caves2002unknown,
  title={Unknown quantum states: the quantum de Finetti representation},
  author={Caves, Carlton M and Fuchs, Christopher A and Schack, R{\"u}diger},
  journal={Journal of Mathematical Physics},
  volume={43},
  number={9},
  pages={4537--4559},
  year={2002},
  publisher={American Institute of Physics}
}

@article{cabello2002bell,
  title={Bell’s theorem with and without inequalities for the three-qubit Greenberger-Horne-Zeilinger and W states},
  author={Cabello, Ad{\'a}n},
  journal={Physical Review A},
  volume={65},
  number={3},
  pages={032108},
  year={2002},
  publisher={APS},
doi={10.1103/PhysRevA.65.032108}
}

@article{dur2000three,
  title={Three qubits can be entangled in two inequivalent ways},
  author={D{\"u}r, Wolfgang and Vidal, Guifre and Cirac, J Ignacio},
  journal={Physical Review A},
  volume={62},
  number={6},
  pages={062314},
  year={2000},
  publisher={APS},
doi={10.1103/PhysRevA.62.062314}
}

@inproceedings{bartschi2019deterministic,
  title={Deterministic preparation of Dicke states},
  author={B{\"a}rtschi, Andreas and Eidenbenz, Stephan},
  booktitle={International Symposium on Fundamentals of Computation Theory},
  pages={126--139},
  year={2019},
  organization={Springer},
howpublished = {\url{https://doi.org/10.1007/978-3-030-25027-0\_9}}
}

@article{kiesel2007experimental,
  title={Experimental observation of four-photon entangled Dicke state with high fidelity},
  author={Kiesel, Nikolai and Schmid, Christian and T{\'o}th, Geza and Solano, Enrique and Weinfurter, Harald},
  journal={Physical review letters},
  volume={98},
  number={6},
  pages={063604},
  year={2007},
  publisher={APS},
doi={10.1103/PhysRevLett.98.063604}
}

@book{schuld2018supervised,
  title={Supervised learning with quantum computers},
  author={Schuld, Maria and Petruccione, Francesco},
  volume={17},
  year={2018},
  publisher={Springer},
howpublished = {\url{https://doi.org/10.1007/978-3-319-96424-9}}
}

@article{harper2015movielens,
  title={The movielens datasets: History and context},
  author={Harper, F Maxwell and Konstan, Joseph A},
  journal={Acm transactions on interactive intelligent systems (tiis)},
  volume={5},
  number={4},
  pages={1--19},
  year={2015},
  publisher={Acm New York, NY, USA},
doi={10.1145/2827872}
}

@article{ding2021quantum,
  title={Quantum-inspired support vector machine},
  author={Ding, Chen and Bao, Tian-Yi and Huang, He-Liang},
  journal={IEEE Transactions on Neural Networks and Learning Systems},
  volume={33},
  number={12},
  pages={7210--7222},
  year={2021},
  publisher={IEEE},
doi={10.1109/TNNLS.2021.3084467}
}

@article{rebentrost2014quantum,
  title={Quantum support vector machine for big data classification},
  author={Rebentrost, Patrick and Mohseni, Masoud and Lloyd, Seth},
  journal={Physical review letters},
  volume={113},
  number={13},
  pages={130503},
  year={2014},
  publisher={APS},
doi={10.1103/PhysRevLett.113.130503}
}

@article{schuld2020circuit,
  title={Circuit-centric quantum classifiers},
  author={Schuld, Maria and Bocharov, Alex and Svore, Krysta M and Wiebe, Nathan},
  journal={Physical Review A},
  volume={101},
  number={3},
  pages={032308},
  year={2020},
  publisher={APS},
doi={10.1103/PhysRevA.101.032308}
}

@article{da2011practical,
  title={Practical characterization of quantum devices without tomography},
  author={da Silva, Marcus P and Landon-Cardinal, Olivier and Poulin, David},
  journal={Physical Review Letters},
  volume={107},
  number={21},
  pages={210404},
  year={2011},
  publisher={APS},
doi={10.1103/PhysRevLett.107.210404}
}

@article{ioffe2015batch,
  title={Batch normalization: Accelerating deep network training by reducing internal covariate shift},
  author={Ioffe, Sergey},
  journal={arXiv preprint arXiv:1502.03167},
  year={2015}
}

@article{park2020theory,
  title={The theory of the quantum kernel-based binary classifier},
  author={Park, Daniel K and Blank, Carsten and Petruccione, Francesco},
  journal={Physics Letters A},
  volume={384},
  number={21},
  pages={126422},
  year={2020},
  publisher={Elsevier},
doi={10.1016/j.physleta.2020.126422}
}

\end{document}